\documentclass[a4paper,12pt]{article}

\usepackage{amsmath}
\usepackage{amssymb}
\usepackage[dvips]{graphicx}
\usepackage[dvips]{psfrag}
\usepackage{bm}

\newcommand {\be}{\begin{equation}}
\newcommand {\ee}{\end{equation}}
\newcommand {\bea}{\begin{eqnarray}}
\newcommand {\eea}{\end{eqnarray}}
\newcommand {\nn}{\nonumber}
\newcommand {\tr}{{\rm tr}}

\newcommand {\der}{\partial}

\newcommand{\cN}{{\cal N}}
\newcommand{\cO}{{\cal O}}

\newcommand{\vev}[1]{\left\langle #1 \right\rangle}

\newcommand{\cD}{{\cal D}}

\def\Xint#1{\mathchoice
{\XXint\displaystyle\textstyle{#1}}%
{\XXint\textstyle\scriptstyle{#1}}%
{\XXint\scriptstyle\scriptscriptstyle{#1}}%
{\XXint\scriptscriptstyle\scriptscriptstyle{#1}}%
\!\int}
\def\XXint#1#2#3{{\setbox0=\hbox{$#1{#2#3}{\int}$}
\vcenter{\hbox{$#2#3$}}\kern-.5\wd0}}

\def\dashint{\Xint-}

\begin{document}
\thispagestyle{empty} \addtocounter{page}{-1}
\begin{flushright}
OIQP-10-06\\
%
\end{flushright} 
\vspace*{1cm}

\begin{center}
{\large \bf  Spontaneous supersymmetry breaking in matrix models \\
from the viewpoints of localization and Nicolai mapping}\\
\vspace*{1.5cm}
Tsunehide Kuroki$^*$ and Fumihiko Sugino$^\dagger$\\
\vskip0.7cm
{}$^*${\it Department of Physics, Rikkyo University, }\\
\vspace*{1mm}
{\it Nishi-Ikebukuro, Tokyo 171-8501, Japan}\\
\vspace*{0.2cm}
{\tt tkuroki@rikkyo.ac.jp}\\
\vskip0.4cm
{}$^\dagger${\it Okayama Institute for Quantum Physics, } \\
\vspace*{1mm}
{\it Kyoyama 1-9-1, Kita-ku, Okayama 700-0015, Japan}\\
\vspace*{0.2cm}
{\tt fumihiko\_sugino@pref.okayama.lg.jp}\\
\end{center}
\vskip1.0cm
\centerline{\bf Abstract}
\vspace*{0.3cm}
{\small 
In the previous work, it was shown that, in supersymmetric (matrix) discretized quantum mechanics, 
inclusion of an external field twisting the boundary condition of fermions enables us 
to discuss spontaneous breaking of supersymmetry (SUSY) in the path-integral formalism 
in a well-defined way. 
In the present work, 
we continue investigating the same systems from the points of view of 
localization and Nicolai mapping. 
The localization is studied by changing of integration variables in the path integral, 
which is applicable whether or not SUSY is explicitly broken. 
We examine in detail 
how the integrand of the partition function with respect to the integral over the auxiliary field
behaves as the auxiliary field vanishes, which clarifies a mechanism of the localization.  
In SUSY matrix models, we obtain a matrix-model generalization of the localization formula. 
In terms of eigenvalues of matrix variables, 
we observe that 
eigenvalues' dynamics is governed by balance of attractive force from the localization and 
repulsive force from the Vandermonde determinant. 
The approach of the Nicolai mapping works 
even in the presence of the external field. 
It enables us to compute the partition function of SUSY matrix models  
for finite $N$ ($N$ is the rank of matrices) with arbitrary superpotential 
at least in the leading nontrivial order of an expansion with respect to the small external field. 
We 
confirm the restoration of SUSY 
in the large-$N$ limit of a SUSY matrix model with a double-well scalar potential 
observed in the previous work.

}
\vspace*{1.1cm}



\newpage

\section{Introduction}
Spontaneous breaking of supersymmetry (SUSY) is one of the most interesting phenomena 
in quantum field theory. Since in general SUSY cannot be broken by radiative corrections at 
the perturbative level, 
its spontaneous breaking requires understanding of 
nonperturbative aspects of quantum field theory~\cite{Witten:1981nf}. 
In particular, recent developments 
in nonperturbative aspects of string theory heavily rely on the presence of SUSY, 
which is however lost in the standard model. 
Thus, in order to deduce predictions to the real world from string theory, 
it is indispensable and definitely important to investigate a mechanism 
of spontaneous SUSY breaking in a nonperturbative framework of strings. 
Since one of the most promising approaches of nonperturbative formulations of string theory 
is provided by large-$N$ matrix models~\cite{Banks:1996vh,Ishibashi:1996xs,Dijkgraaf:1997vv} 
($N$ is the rank of matrix variables), it will be desirable to understand how SUSY can be spontaneously broken 
in the large-$N$ limit of simple matrix models as a first step. 
For example, in IIB matrix model~\cite{Ishibashi:1996xs} it has been suggested 
that the rotational $SO(10)$ symmetry is spontaneously broken 
in the large-$N$ limit~\cite{Nishimura:2001sx}.  
This tempts us to expect that SUSY is also broken in the large-$N$ limit of this model. 
Analysis of SUSY breaking in simple matrix models would help us find 
a mechanism which is responsible for possible spontaneous SUSY breaking 
in nonperturbative string theory. 

For this purpose, it is desirable to treat systems in which 
spontaneous SUSY breaking takes place in the path-integral formalism, 
because matrix models are usually defined by the path integrals, namely integrals over matrix variables. 
In particular, IIB matrix model~\cite{Ishibashi:1996xs} defined in zero dimension  
can be formulated only by the path-integral formalism. 
Motivated by this, in the previous work~\cite{Kuroki:2009yg}, we constructed the path-integral formalism 
for SUSY (matrix) quantum mechanics on discretized Euclidean time $t\in \{1,2,\cdots, T\}$, 
which includes cases that SUSY is spontaneously broken~\footnote{
Notice that SUSY can be spontaneously broken in systems defined in less than one-dimension as discussed 
in~\cite{Kuroki:2009yg}. Namely, an analog of the Mermin-Wagner-Coleman theorem does not hold for SUSY.}. 
It is formulated in a well-defined way, 
by introducing an external field, which explicitly breaks the SUSY, to twist the boundary condition of fermions 
in the Euclidean time direction. 
In this setup, we compute an order parameter of SUSY breaking such as the expectation value 
of an auxiliary field in the presence of the external field. If it remains nonvanishing 
after turning off the external field, it shows that SUSY is spontaneously broken 
because it implies that the effect of the infinitesimal external field we have introduced 
at the beginning remains. 
Here, it should be noticed that, if we are interested in the large-$N$ limit, 
we have to take it before turning off the external field, 
which is reminiscent of the thermodynamic limit of the Ising model 
taken before turning off the magnetic field 
in detecting the spontaneous ${\bf Z}_2$ breaking. 
In the formalism proposed in~\cite{Kuroki:2009yg}, for the expectation value 
of an auxiliary field, the external field plays the role of a regulator  
by which it is computed in a well-defined manner. 
In particular, if we take the periodic boundary condition for fermions, the partition function 
is essentially the Witten index~\cite{Witten:1982df} 
which vanishes when the SUSY is spontaneously broken. 
However, since the external field explicitly breaks the SUSY by a small amount, 
the partition function with the external field becomes nonzero, 
and the expectation value normalized by the partition function is well-defined. 
Moreover, we have seen that 
the expectation value of the auxiliary field is also well-defined in the limit turning off 
the external field due to cancellation of its dependence between the numerator 
and the denominator. This shows how the expectation value of the auxiliary field can 
have nonzero value in the path-integral formalism. 

In view of this, it is quite important 
to calculate the partition function in the presence of the external field 
in the path integral for systems 
which spontaneously break SUSY. 
Especially it would be better to calculate it in matrix models at finite $N$ 
in order to observe breaking/restoration of SUSY in the large-$N$ limit. 
In this paper, we address this problem by utilizing two methods: 
localization~\cite{Witten:1991mk} and Nicolai mapping~\cite{Nicolai:1979nr}. As for the localization, 
we make change of integration variables in the path integral, which is always possible 
whether or not the SUSY is explicitly broken (the external field is on or off). 
It is investigated in detail how the integrand of the partition function with respect to the integral over 
the auxiliary field behaves as the auxiliary field approaches to zero. 
It plays a crucial role to understand the localization from the change of variables. 
To our knowledge, this kind of investigation has not been found in the literature.  
In the case of discretized SUSY quantum mechanics with $Q$-SUSY preserved, it implies that 
the path integral receives contributions only from the fixed points of $Q$-transformation 
and reproduces known results for the localization formula. 
In particular, for the $T=1$ case corresponding to the zero-dimensional model, 
the fixed points of $Q$-transformation are nothing but the critical points of superpotential, 
i.e. zeros of the first derivative of superpotential.  
In the case of SUSY matrix models, analogous localization formula can be obtained. 
However, in terms of eigenvalues of matrix variables, an interesting phenomenon occurs. 
Localization attracts the eigenvalues to the critical points of superpotential, while the square of 
the Vandermonde determinant arising from the measure factor prevents the eigenvalues from collapsing. 
The dynamics of the eigenvalues is governed by balance of attractive force from the localization 
and repulsive force from the Vandermonde determinant. 
Without the external field, contribution to the partition function 
from each eigenvalue distributed around some critical point is derived for a general superpotential. 
In the case of a double-well scalar potential, it leads 
to the statement (4.17) in~\cite{Kuroki:2009yg} in the large-$N$ limit.  
When the external field is turned on, computation is still possible, but we find that 
a method by the Nicolai mapping is more effective. 
Interestingly, it works for SUSY matrix models  
even in the presence of the external field which explicitly breaks SUSY. 
It enables us to calculate the partition function at least in the leading nontrivial order of an expansion 
with respect to the small external field 
for finite $N$. We can take the large-$N$ limit of our result before turning off 
the external field and detect whether SUSY is spontaneously broken or not in the large-$N$ limit. 
As a byproduct of the analysis, we give a clear argument for the restoration of SUSY
in a SUSY matrix model with a double-well scalar potential 
at large $N$, which was observed in \cite{Kuroki:2009yg}. 

This paper is organized as follows. 
In the next section, we consider change of variables 
in the path integral for discretized SUSY quantum mechanics leading to localization. 
It is pointed out how it works by investigating the behavior of the integrand of the partition function 
as the auxiliary field becomes small. 
In section~\ref{sec:loc_in_MQM}, 
a similar method is applied to SUSY matrix models, and a matrix-model generalization 
of the localization formula is derived. 
In section~\ref{sec:Nicolai}, 
we make an expansion of the partition function with respect to a small external field 
and derive a formula for a general superpotential  
in the leading nontrivial order of the expansion. 
It is valid for arbitrary $N$. 
By applying it to the case of a double-well scalar potential, 
we confirm the restoration of SUSY in the large-$N$ limit of this model discussed 
in \cite{Kuroki:2009yg}. 
We summarize the result so far and discuss future directions in section~\ref{sec:summary}. 
Details of localization in discretized SUSY quantum mechanics with $T\geq 2$ are discussed 
in appendix~\ref{app:loc_generalT}. 
Finally, some computational details are presented in appendix~\ref{app:Y_N}.

\section{Change of variables and localization in discretized SUSY quantum mechanics} 
\label{sec:loc}
\setcounter{equation}{0}
As discussed in \cite{Kuroki:2009yg}, in order to discuss
spontaneous SUSY breaking in the path-integral formalism of (discretized) SUSY quantum mechanics or 
SUSY matrix models, 
we introduce an external field to twist the boundary condition of fermions 
in the Euclidean time direction and observe whether an order parameter of SUSY 
breaking remains nonzero after turning off the external field. 
This motivates us to calculate the partition function in the presence of the external field. 
In the following, we consider systems of SUSY quantum mechanics on discretized Euclidean time 
$t\in \{1,2,\cdots,T\}$. 
As shown below, it is possible to introduce such an external field even in zero dimension ($T=1$). 
Therefore, by considering the simplest zero-dimensional models, it is expected that 
we can extract some essential properties of the partition function in the presence of the external field 
without touching technical complexity. 
In this section, we consider change of variables which leads to the localization of contribution to 
the path integral 
and will be useful in the computation of the partition function. 

\subsection{Introduction of external field}
\label{subsec:alpha}
We begin with SUSY quantum mechanics whose action is given by 
\begin{align}
S=\int_0^\beta dt\,  \left[\frac12 B^2 +iB\left(\dot{\phi} + W'(\phi)\right) 
+\bar{\psi}\left(\dot{\psi} +W''(\phi)\psi\right)\right], 
\label{S_SUSYQM}
\end{align}
where the Euclidean time direction is compactified by $\beta$, ($\dot{\mbox{ }}$) means the time derivative, 
and $W'(\phi)$ and $W''(\phi)$ are the first and second derivatives of the superpotential $W(\phi)$ with respect to $\phi$. 
In this paper, we focus on the case that $W(\phi)$ is a polynomial of $\phi$. 
$S$ is invariant under one-dimensional 
$\cN=2$ SUSY transformations generated by $Q$ and $\bar{Q}$, which act on fields as  
\begin{align}
Q\phi =\psi,~~~Q\psi=0,~~~Q\bar{\psi} =-iB,~~~QB=0, 
\label{QSUSY}
\end{align}
and 
\begin{align}
\bar{Q} \phi = -\bar{\psi},~~~\bar{Q}\bar{\psi} = 0,~~~
\bar{Q} \psi = -iB +2\dot{\phi},~~~\bar{Q} B = 2i\dot{\bar{\psi}}. 
\label{Qbar_SUSY}
\end{align}
They satisfy the algebra 
\be
Q^2=\bar{Q}^2 =0, \qquad \{Q, \bar{Q}\}=2\partial_t.
\ee 
The invariance of $S$ follows from its $Q$- or $Q\bar Q$-exactness: 
\begin{align}
S & = Q\int_0^\beta dt \, \bar{\psi}\left\{ \frac{i}{2} B -\left(\dot{\phi} +W'(\phi)\right)\right\} 
= Q\bar{Q} \int_0^\beta dt \, \left( \frac12\bar{\psi}\psi +W(\phi)\right). 
\label{S_SUSYQM2}
\end{align}
The partition function is defined by 
\begin{align}
Z=\int \cD B \cD\phi \cD\psi \cD\bar\psi \,e^{-S}
\label{Z}
\end{align}
with the path-integral measure normalized as
\be
\int \cD\phi\,e^{-\int_0^\beta dt\,\frac12 \phi(t)^2}=\int\cD B\,e^{-\int_0^\beta dt\,\frac12 B(t)^2}=1, 
\qquad 
\int\cD\psi\cD\bar\psi\, e^{-\int_0^\beta dt\,\bar\psi(t)\psi(t)}=1.
\ee
It is pointed out in \cite{Kuroki:2009yg} that if we take the periodic boundary condition 
for all fields, \eqref{Z} is equivalent to the Witten index~\cite{Witten:1982df}, 
which vanishes when SUSY is spontaneously broken. It means that 
the expectation value normalized by the partition function is generally ill-defined in such a case. 
Since the vanishing partition function originates from cancellation between 
bosonic and fermionic states, we will introduce an external field 
which explicitly breaks the SUSY, in order to resolve the degeneracy and to fix a single vacuum 
in which the SUSY is broken. It is analogous to the magnetic field introduced in the Ising model 
in detecting the ${\bf Z}_2$ symmetry breaking.  
Let us modify the periodic boundary condition of the fermions to 
a twisted one as 
\begin{align}
\psi(t+\beta)=e^{i\alpha}\psi(t),~~~\bar\psi(t+\beta)=e^{-i\alpha}\bar\psi(t).
\end{align}
Here, the twist $\alpha$ corresponds to the external field. 
It is shown that in the presence of $\alpha$, \eqref{Z} does not vanish and the normalized expectation value 
of the auxiliary field $\vev{B}_\alpha$ is well-defined. Moreover, it turns out that $\vev{B}_\alpha$ does not 
depend on $\alpha$ and therefore $\alpha\rightarrow 0$ limit is also well-defined 
\cite{Kuroki:2009yg}. In this sense, $\alpha$ plays the role 
of a regulator by which we can calculate the expectation value of an order parameter of SUSY breaking 
unambiguously. 
Thus, the external field $\alpha$ provides a framework for discussing spontaneous SUSY breaking 
in the path-integral formalism.  

As a discretized version of \eqref{S_SUSYQM2} and (\ref{S_SUSYQM}), we consider 
\bea
S 
& = & Q \sum_{t=1}^T \bar{\psi}(t) \left\{ \frac{i}{2} B(t) 
 -\left(\phi(t+1) -\phi(t) +W'(\phi(t))\right)\right\} 
\nonumber \\
 & = & \sum_{t=1}^T \left[\frac12 B(t)^2 +iB(t)\left\{\phi(t+1)-\phi(t)+W'(\phi(t))\right\} \right. \nn \\
 &  &\hspace{2cm} 
\left. \frac{}{} +\bar{\psi}(t) \left\{\psi(t+1) -\psi(t) +W''(\phi(t)) \psi(t)\right\} \right],  
 \label{dQM_S2}
\eea
which preserves $Q$-SUSY but breaks $\bar{Q}$-SUSY 
by the discretization~\footnote{When $T=1$ and all the variables obey the periodic boundary condition, 
the action is nothing but the dimensional reduction of (\ref{S_SUSYQM}) and 
invariant under both $Q$ and $\bar{Q}$ as seen in the next subsection.}. 
Let us express by $S_\alpha$ the action (\ref{dQM_S2}) under the twisted boundary condition 
\be
\phi(T+1)=\phi(1), \qquad \psi(T+1) = e^{i\alpha} \psi(1). 
\label{TBC_alpha}
\ee
Namely, $\phi(T+1)$ and $\psi(T+1)$ appearing in (\ref{dQM_S2}) are understood to be replaced with  
$\phi(1)$ and $e^{i\alpha}\psi(1)$, respectively. 
Then the partition function is defined as 
\be
Z_\alpha = \left(\frac{-1}{2\pi}\right)^T
\int \prod_{t=1}^T\left(dB(t)\,d\phi(t)\,d\psi(t)\,d\bar{\psi}(t)\right) 
\, e^{-S_\alpha}. 
\label{Z_alpha_T}
\ee
We will fix the sign convention of integrals over Grassmann numbers as 
\be
\int d\psi(t) \,\psi(t') = \int d\bar\psi(t)\,\bar\psi(t') = \delta_{t, t'}. 
\ee

In the simplest case $T=1$, the action and the partition function are expressed as  
\begin{align}
S_\alpha & =\frac12 B^2 +iB W'(\phi) +\bar{\psi} \left(e^{i\alpha}-1 +W''(\phi)\right) \psi, 
\nonumber \\
Z_\alpha & =-\frac{1}{2\pi}\int dB \,d\phi\, d\psi \,d\bar{\psi} \, e^{-S_\alpha}.
\label{twistedS&Z}
\end{align}
We see that the effect of the external field remains even in the zero-dimensional model 
and breaks the SUSY.

\subsection{Localization in $T=1$ discretized SUSY quantum mechanics}
\label{subsec:Loc_in_QM}
As a simple example of localization, we first discuss 
the $T=1$ system (\ref{twistedS&Z}) under the periodic boundary condition ($\alpha=0$): 
\begin{align}
S_0 & =\frac12 B^2 +iB W'(\phi) +\bar{\psi} W''(\phi) \psi, 
\nonumber \\
Z_0 & =-\frac{1}{2\pi}\int dB \,d\phi \,d\psi \,d\bar{\psi} \, e^{-S_0}.
\label{untwistedS&Z}
\end{align}
$S_0$ preserves the ${\cal N}=2$ SUSY 
\begin{align}
Q\phi =\psi,~~~Q\psi=0,~~~Q\bar{\psi} =-iB,~~~QB=0, 
\label{QSUSY0dim}
\end{align}
and 
\begin{align}
\bar{Q} \phi = -\bar{\psi},~~~\bar{Q}\bar{\psi} = 0,~~~
\bar{Q} \psi = -iB,~~~\bar{Q} B =0, 
\label{Qbar_SUSY0dim}
\end{align}
which are reduction of \eqref{QSUSY} and \eqref{Qbar_SUSY} to zero dimension. 

Let us consider the following field redefinition~\footnote{
The argument leading to localization from a field redefinition is based on 
Chapter 9.3 in~\cite{Hori:2003ic}. However, the auxiliary field is not introduced there 
and the treatment of path-integral measure seems somewhat incomplete. 
For instance, the second term of (9.35) in~\cite{Hori:2003ic} does not vanish in general, 
contrary to the claim in~\cite{Hori:2003ic}\label{footnote}.  
}
$(B,\phi,\psi,\bar\psi) \rightarrow (B,\tilde\phi,\psi,\bar\epsilon)$: 
\begin{align}
\phi=\tilde\phi+\bar\epsilon\psi,~~~\bar\psi=-i\bar\epsilon B.
\label{fieldredef}
\end{align}
Note that, from the SUSY transformation \eqref{QSUSY0dim}, 
these can be rewritten as~\footnote{
We can also consider another field redefinition which is expressed as $\bar{Q}$ transformation, 
and the argument proceeds similarly.   
}    
\begin{align}
\phi=\tilde\phi+\bar\epsilon Q\tilde\phi,~~~\bar\psi=0+\bar\epsilon Q\bar\psi.
\label{SUSYtransf}
\end{align}
It implies that $\phi$ and $\bar\psi$ are expressed 
as the SUSY transformation from $\tilde\phi$ and $\bar{\psi}=0$ respectively, 
and that the SUSY transformation parameter $\bar{\epsilon}$ is regarded 
as a fermionic variable instead of $\bar{\psi}$. 
Then, from the SUSY invariance of $S_0$, 
\begin{align}
S_0(B,\phi,\psi,\bar\psi)
&=S_0(B+\bar\epsilon QB,\tilde\phi+\bar\epsilon Q\tilde\phi,
     \psi+\bar\epsilon Q\psi,0+\bar\epsilon Q\bar\psi) \nonumber \\
&=S_0(B,\tilde\phi,\psi,\bar\psi=0)=\frac12B^2+iBW'(\tilde\phi),
\label{newS0}
\end{align}
which is independent of $\bar\epsilon$. 
(This expression can be directly derived by using $\tilde\phi=\phi-i\bar\psi\psi/B$ 
obtained from \eqref{fieldredef}.)  
Furthermore, since the Jacobian associated with 
\eqref{fieldredef} is computed as 
\begin{align}
dB\,d\phi\, d\psi \,d\bar\psi=\frac{i}{B}\,dB\,d\tilde\phi \,d\psi \,d\bar\epsilon,
\label{QMJacobian}
\end{align}
and the $B$-integral in the partition function looks singular at $B=0$, 
we can say that the change of variables \eqref{fieldredef} is always possible 
for $B\neq 0$. However, notice that, 
if other $B$-dependence than the Jacobian \eqref{QMJacobian} arises 
which makes the $B$-integral nonsingular at the origin, 
(\ref{fieldredef}) is possible even at $B=0$. 
We will see such an example explicitly below. 

By using (\ref{newS0}) and (\ref{QMJacobian}), we find that the path integral 
of the partition function $Z_0$ given in \eqref{untwistedS&Z} is localized at 
$B=0$. Namely, if we divide the integration region of $B$ in \eqref{untwistedS&Z} 
into the vicinity of $B=0$: $\{B\,|\,|B|<\varepsilon\}$ and its complement 
$\{B\,|\, |B|\ge\varepsilon\}$ as
\begin{align}
Z_0=Z^{(0)}_0+\tilde Z_0,
\end{align}
where 
\begin{align}
&Z^{(0)}_0=\int_{|B|<\varepsilon}dB \,\Xi_0(B), \qquad 
\tilde Z_0=\int_{|B|\ge\varepsilon}dB \,\Xi_0(B), \nn \\
&\Xi_0(B) \equiv -\frac{1}{2\pi}\int d\phi \,d\psi \,d\bar\psi\, e^{-S_0}
\label{decomposition}
\end{align}
with $0<\varepsilon \ll 1$, 
then $\tilde Z_0$ is shown to vanish due to the trivial $\bar\epsilon$-integral after the above change of variables. 

On the other hand, for the purpose of examining whether \eqref{fieldredef} works 
even in computation of $Z^{(0)}_0$, we have to take account of  other $B$-dependence 
than the one in \eqref{QMJacobian} and to observe if $B$-integral still diverges or not. 
In order to see the behavior of $\Xi_0(B)$ in the vicinity of $B=0$, 
it is instructive to try the change of variables for $|B|<\varepsilon$. 
We have 
\be
\Xi_0(B) = -\frac{1}{2\pi}\,\frac{i}{B}\,e^{-\frac12B^2}
\int^\infty_{-\infty} d\tilde{\phi}\,e^{-iBW'(\tilde{\phi})}\,\int d\psi\,d\bar\epsilon.
\label{Xi_0(B)_fieldredef0}
\ee
When $W'(\phi)$ is a polynomial of degree $p$ ($p\ge 2$): 
\be
W'(\phi)=g_p\phi^p + g_{p-1}\phi^{p-1}+ \cdots + g_0, 
\label{W'}
\ee
we rescale as 
\be
\tilde{\phi} = |B|^{-\frac{1}{p}}\phi'
\label{rescale_phi}
\ee 
to extract $|B|$-dependence from 
the $\tilde{\phi}$-integral :  
\be
\int^\infty_{-\infty} d\tilde{\phi}\,e^{-iBW'(\tilde{\phi})}= 
\frac{1}{|B|^{\frac{1}{p}}} \int^\infty_{-\infty} d\phi'\,e^{-i\,{\rm sgn}(B)\,g_p\phi'^p}
\left[1+\cO(\varepsilon^{1/p})\right] .
\ee
Note that, since $|B|<\varepsilon$, the first term of $W'(\phi)$ in \eqref{W'} 
becomes the most important after the rescaling \eqref{rescale_phi}. 
Hence, we see that $\Xi_0(B)$ is singular as $|B|^{-1-\frac{1}{p}}$ near the origin,  
and $Z_0^{(0)}$ can be expressed as~\footnote{
The $\phi'$-integral is computed as 
\be
\int^\infty_{-\infty} d\phi'\,\sin\left(g_p\phi'^p\right)= \begin{cases} 0 & (p: \mbox{odd}) \\
\frac{{\rm sgn}(g_p)}{|g_p|^{\frac{1}{p}}}\, 2\sin\left(\frac{\pi}{2p}\right)\,
\Gamma\left(1+\frac{1}{p}\right) & (p: \mbox{even}). \end{cases}
\ee
}
\bea
Z_0^{(0)} & = & 
-\frac{1}{\pi}\,
\left(\int^\varepsilon_0 dB\,\frac{1}{B^{1+\frac{1}{p}}}\,e^{-\frac12B^2} \right)
\left(\int^\infty_{-\infty}d\phi'\,\sin\left(g_p\phi'^p\right)\right)\,
\int d\psi\int d\bar\epsilon \nn \\
& & \times\left[1+\cO(\varepsilon^{1/p})\right] .
\label{Z_0^0_fieldredef}
\eea
The integrals of $\psi$ and $\bar\epsilon$ vanish, while the $B$-integral is divergent.  
Since the expression (\ref{Z_0^0_fieldredef}) is of indefinite form $\infty\times 0$, 
it is found that the change of variables (\ref{fieldredef}) is not appropriate to compute $Z_0^{(0)}$. 
(For $p=1$ case, because the $\tilde\phi$-integral in (\ref{Xi_0(B)_fieldredef0}) gives $\delta(B)$, 
it is clear that the $B$-integral in $Z_0^{(0)}$ is divergent.) 
The indefinite form of $Z_0^{(0)}$ under the change of variables (\ref{fieldredef}) 
can be understood to reflect that 
$Z_0^{(0)}$ possibly takes a nonzero value if it is evaluated in a well-defined manner.   

\paragraph{Unnormalized expectation values}
For the unnormalized expectation values of $B^n$ ($n\ge 1$): 
\be
\vev{B^n}' \equiv -\frac{1}{2\pi}\int dB\,d\phi\,d\psi\,d\bar\psi\,B^n \,e^{-S_0}, 
\ee
we use the same change of variables to have 
\be
\vev{B^n}' = \frac{-i}{2\pi}\int dB\,d\tilde{\phi}\,d\psi\,d\bar{\epsilon}\,B^{n-1}\,
e^{-\frac12B^2-iBW'(\tilde{\phi})}.
\ee
In contrast to the case of the partition function, we will see that the change of variables (\ref{fieldredef})  
is possible for any value of $B$ in evaluating $\vev{B^n}'$.

Notice that the $B$-integral is not singular at $B=0$ for $n\ge 1$. 
In fact, in the region $|B|<\varepsilon$, the $B$-integral after the rescaling (\ref{rescale_phi}) 
gives a finite value:  
\be
\int^\varepsilon_0 dB\,B^{n-1-\frac{1}{p}}\,e^{-\frac12 B^2} = \frac{\varepsilon^{n-\frac{1}{p}}}{n-\frac{1}{p}}
\left(1+\cO(\varepsilon^2)\right) 
\ee  
for $p\ge 2$ case of $W'(\phi)$ in (\ref{W'}). 
In $p=1$ case, the $B$-integral is clearly finite as 
\be
\int^\varepsilon_{-\varepsilon}dB\,B^{n-1}\,\delta(B)=\delta_{n, 1}.
\label{Bn_unVEV_p=1}
\ee
It indicates that the change of variables (\ref{fieldredef}) is allowed for any value of $B$.
  
Thus, for all $p\geq 1$ in $W'(\phi)$, we obtain  
\be
\vev{B^n}' = 0 \qquad (n\ge 1)
\label{Bn_unVEV}
\ee 
from the trivial $\bar\epsilon$-integral: $\int d\bar\epsilon =0$.  

(\ref{Bn_unVEV}) implies that the localization to $B=0$ is realized 
in such a way that $\Xi_0(B)$ is proportional to $\delta(B)$ 
(without derivatives of $\delta(B)$). 
It can be directly derived as follows. 
Applying the Nicolai mapping $X=W'(\phi)$ to 
\be
\Xi_0(B)=\frac{1}{2\pi}\int d\phi\,e^{-\frac12B^2-iBW'(\phi)}\,W''(\phi),
\ee
we have 
\be
\Xi_0(B)=e^{-\frac12B^2}\,\frac{1}{2\pi}\,\sharp\,\int^\infty_{-\infty}dX\,e^{-iBX} =\sharp\,\delta(B),
\label{Xi0_Nicolaimapping}
\ee
where $\sharp$ is the mapping degree of $X=W'(\phi)$. For $W'(\phi)$ given as 
a polynomial (\ref{W'}) with the degree $p$, 
\be
\sharp = \begin{cases} {\rm sgn}(g_p) & \mbox{for $p$: odd} \\
                          0           & \mbox{for $p$: even}. \end{cases}
\label{mapping_degree}                           
\ee

\paragraph{Localization to $W'(\phi)=0$}
Since 
\be
\vev{e^{-\frac{u-1}{2}B^2}}' = \sum_{n=0}^\infty \frac{1}{n!}\left(-\frac{u-1}{2}\right)^n \vev{B^{2n}}' 
= \vev{1}' = Z_0,
\ee
which follows from (\ref{Bn_unVEV}) for an arbitrary parameter $u$, the partition function can be expressed as 
\be
Z_0  = \frac{-1}{2\pi}\int dB\,d\phi \,d\psi \,d\bar\psi\, 
e^{-\frac{u}{2}B^2-iBW'(\phi)-\bar{\psi}W''(\phi)\psi}.  
\ee
Note that $Z_0$ does not depend on the value of $u$. 
Let us take $u>0$ to perform the $B$-integration first. Then,  
\bea
Z_0 & = & -\int d\phi\,d\psi\,d\bar\psi\,\frac{1}{\sqrt{2\pi u}} \,e^{-\frac{1}{2u} W'(\phi)^2} \,
e^{-\bar\psi W''(\phi)\psi} \nn \\
& = & \int d\phi\,\frac{1}{\sqrt{2\pi u}} \,e^{-\frac{1}{2u} W'(\phi)^2} \,W''(\phi).
\label{Z_0_T=1_loc}
\eea
In the limit $u\to 0$, the factor $\frac{1}{\sqrt{2\pi u}} \,e^{-\frac{1}{2u} W'(\phi)^2}$ becomes  
$\delta(W'(\phi))$, which directly leads to 
localization to the critical points of the superpotential satisfying $W'(\phi)=0$.   

In the case that the superpotential is a polynomial and its critical points are nondegenerate 
(i.e. $W'=0$, $W''\neq 0$ at the critical points), the limit $u\to 0$ yields 
a well-known formula of the localization: 
\be
Z_0 = \int^\infty_{-\infty} d\phi\,\delta(W'(\phi))\,W''(\phi) 
= \sum_{\phi:\,W'(\phi)=0} \frac{W''(\phi)}{|W''(\phi)|},
\label{Z_0_T=1_loc_2}
\ee 
where the sum is taken over the critical points. 
Comparing (\ref{Z_0_T=1_loc_2}) with the $B$-integral of (\ref{Xi0_Nicolaimapping}), we obtain 
\be
\sharp = \sum_{\phi:\,W'(\phi)=0} \frac{W''(\phi)}{|W''(\phi)|}. 
\label{mapping_degree_localization}
\ee
The same result can be obtained by the one-loop computation around the critical points. 
Let $\phi_c$ be a critical point of $W'(\phi)$ and $\varphi$ be a fluctuation around $\phi_c$: 
\be
\phi=\phi_c+\sqrt{u}\,\varphi.
\ee
Then, (\ref{Z_0_T=1_loc}) becomes 
\be
Z_0 = \sum_{\phi_c:\,W'(\phi_c)=0} \frac{1}{\sqrt{2\pi}}\,\int^\infty_{-\infty} d\varphi\,
e^{-\frac12 W''(\phi_c)^2\varphi^2}\,W''(\phi_c) + \cO(\sqrt{u}),   
\label{Z_0_T=1_loc_3}
\ee
where contribution around each of critical points has to be summed if $W'(\phi)$ has two or more critical 
points. 
It is easy to see that the one-loop computation of $\varphi$ reproduces the RHS of (\ref{Z_0_T=1_loc_2}). 
Note that higher loop contributions are $\cO(\sqrt{u})$ and negligible in the $u\to 0$ limit.  
Thus, the one-loop computation around the critical points of the superpotential gives 
the exact answer of the partition function in all order of perturbation theory.

\subsection{Localization in the presence of external field}
\label{subsec:Loc_in_QM2}
Next, we consider the system \eqref{twistedS&Z} with the twisted boundary condition 
($\alpha\neq 0$).  
The field redefinition \eqref{fieldredef} 
changes $S_\alpha$ to 
\begin{align}
S_\alpha=\frac12B^2+iBW'(\tilde\phi)-i(e^{i\alpha}-1)\bar\epsilon B\psi, 
\end{align} 
which has $\bar\epsilon$-dependence due to the twist $\alpha$. 
We again separate the integration region of $B$ 
as  
\begin{align}
&Z_\alpha=Z_\alpha^{(0)}+\tilde Z_\alpha, \nonumber \\
&Z_\alpha^{(0)}=\int_{-\varepsilon}^\varepsilon dB\,\Xi_\alpha(B), \qquad 
\tilde Z_\alpha=\int_{|B|\ge\varepsilon}dB \,\Xi_\alpha(B), \nn \\
&\Xi_\alpha(B)\equiv -\frac{1}{2\pi}\,\int d\phi \,d\psi \,d\bar\psi\,
e^{-S_\alpha}. 
\label{separation1}
\end{align}
For $B\neq 0$, the change of variables \eqref{fieldredef} leads $\Xi_\alpha(B)$ to~\footnote{\label{fn:T=1}
Note that the second line of (\ref{Ztildewithalpha}) is not valid for $B\sim 0$. 
In the expansion of the last factor 
$e^{i(e^{i\alpha}-1)\bar{\epsilon}B\psi}= 1+ i(e^{i\alpha}-1)\bar{\epsilon}B\psi$ in the first line of 
(\ref{Ztildewithalpha}), 
we should not drop the first term ``1" although it yields vanishing Grassmann integrals. 
The reason is that the $B$-integral in $Z^{(0)}_\alpha$ is singular and that the total contribution 
to $Z^{(0)}_\alpha$ is of an indefinite form $\infty \times 0$ 
which cannot be simply regarded as zero. 
Since the corresponding term is nothing but $Z_0^{(0)}$, we find  
\be
Z_\alpha^{(0)}= Z_0^{(0)} + (e^{i\alpha}-1)\int^\varepsilon_{-\varepsilon}dB \int d\tilde{\phi}\, 
 e^{-\frac12 B^2-iBW'(\tilde{\phi})}.
\ee
}
\begin{align}
\Xi_\alpha(B)&=-\frac{1}{2\pi}\,\frac iB\int d\tilde\phi \,d\psi \,d\bar\epsilon\,
e^{-\frac12B^2-iBW'(\tilde\phi)}\,e^{i(e^{i\alpha}-1)\bar\epsilon B\psi} \nonumber \\
&=(e^{i\alpha}-1)\,\frac{1}{2\pi}\int d\tilde\phi\,e^{-\frac12B^2-iBW'(\tilde\phi)}. 
\label{Ztildewithalpha}
\end{align}
$\Xi_\alpha(B)$ and thus $\tilde Z_\alpha$ do not vanish in general by the effect of the twist $e^{i\alpha}-1$. This suggests that the localization is slightly violated by the twist. 
In an exceptional case of $W'(\phi)$ being linear, $\tilde{Z}_\alpha$ vanishes, because 
the $\tilde{\phi}$-integral gives $\delta(B)$ whose support is out of the integration region of $B$.  
Thus, the localization persists in the presence of the twist $\alpha$ when $W'(\phi)$ is linear.  

On the other hand, the twisted partition function $Z_\alpha$ is computed 
without using the change of variables (\ref{fieldredef}) as 
\bea
Z_\alpha & = & \frac{1}{2\pi} \int dB\,d\phi\,e^{-\frac12 B^2 -iBW'(\phi)}\,\left(e^{i\alpha}-1+W''(\phi)\right)
\nn \\
 & = & Z_0 + (e^{i\alpha}-1)\,\frac{1}{\sqrt{2\pi}} \int d\phi\,e^{-\frac12 W'(\phi)^2}.  
\label{Z_alpha} 
\eea
The second term represents the effect of the twist, which is the sum of $\tilde{Z}_\alpha$ 
and the effect of the twist on $Z_\alpha^{(0)}$ (i.e. $Z_\alpha^{(0)}-Z_0^{(0)}$). 
Note that $Z_0=Z_0^{(0)}$ from the localization seen in the previous subsection. 
The second term in \eqref{Z_alpha} of course tends to zero in the $\alpha\rightarrow 0$ limit, 
but notice that it becomes relevant when the SUSY is spontaneously broken, namely $Z_0=0$. 
Let us take a closer look at the effect of the twist on $Z_\alpha^{(0)}$ 
which is the contribution from the vicinity of $B=0$. 
When $W'(\phi)$ is linear ($W'(\phi)=g_1\phi+g_0$), 
\bea
Z_\alpha^{(0)}-Z_0^{(0)} & = & (e^{i\alpha}-1)\,\frac{1}{2\pi}\int^\varepsilon_{-\varepsilon} dB\,e^{-\frac12 B^2}
\int^\infty_{-\infty} d\phi\,e^{-iB(g_1\phi+g_0)} \nn \\
& = & \frac{e^{i\alpha}-1}{|g_1|}
\eea
is not zero even in the $\varepsilon \to 0$ limit, while a similar calculation tells us that 
$\tilde Z_\alpha=0$. 
In contrast, when $W'(\phi)$ is a polynomial (\ref{W'}) with the degree $p\geq 2$, 
we can show that $Z_\alpha^{(0)}-Z_0^{(0)}$ vanishes as $\varepsilon \to 0$.   
Similarly to the argument in the previous subsection, rescaling 
$\phi \to |B|^{-\frac{1}{p}}\,\phi'$ yields~\footnote{
The $\phi'$-integral is calculated as 
\be
\int^\infty_{-\infty}d\phi'\,\cos(g_p\phi'^p)= \frac{1}{|g_p|^{\frac{1}{p}}}\,2\cos\left(\frac{\pi}{2p}\right) 
\Gamma\left(1+\frac{1}{p}\right).
\ee
}  
\bea
Z_\alpha^{(0)}-Z_0^{(0)} & = & (e^{i\alpha}-1)\,\frac{1}{2\pi}\int^\varepsilon_{-\varepsilon} dB\,
\frac{1}{|B|^{\frac{1}{p}}}\,e^{-\frac12 B^2} 
\int d\phi'\,e^{-i\,{\rm sgn} (B) g_p\phi'^p} \left[1+\cO(\varepsilon^{1/p})\right] \nn \\
 & = & (e^{i\alpha}-1)\,\frac{1}{\pi}\left(\int^\varepsilon_0 dB\,
\frac{1}{B^{\frac{1}{p}}}\,e^{-\frac12 B^2}\right) \int^\infty_{-\infty}d\phi'\,\cos(g_p\phi'^p)  \nn \\
 & & \times \left[1+\cO(\varepsilon^{1/p})\right].
\eea
Here the $B$-integral is not singular, and we find  
\be
Z_\alpha^{(0)} -Z_0^{(0)} = \cO\left(\varepsilon^{1-\frac{1}{p}}\right) \to 0 \qquad (\varepsilon \to 0)
\ee
for $p\geq 2$. 
Thus, we conclude that the effect of the external field in the partition function is irrelevant in the 
vicinity of $B=0$ except the case that $W'(\phi)$ is linear.  
Note that, when $W'(\phi)$ is linear, the fermion determinant does not contain field variables, and that 
$\Xi_\alpha(B)$ is proportional to $\Xi_0(B)$ as $
\Xi_\alpha(B) = \frac{e^{i\alpha}-1+g_1}{g_1}\,\Xi_0(B)$. 
This explains the persistence of the localization under the twist.

\paragraph{Unnormalized expectation values} 
For the unnormalized expectation values of $B^n$ ($n\geq 1$): 
\be
\vev{B^n}_\alpha'\equiv -\frac{1}{2\pi}\int dB\,d\phi\,d\psi\,d\bar\psi\,B^n \,e^{-S_\alpha}, 
\ee
the change of variables (\ref{fieldredef}) leads to 
\be
\vev{B^n}_\alpha' = (e^{i\alpha}-1)\,\frac{1}{2\pi}\int dB\,d\tilde{\phi}\,B^n\,
e^{-\frac12 B^2-iBW'(\tilde{\phi})}. 
\ee
Note that, since the $B$-integral is not singular, the change of variables is always possible.    
Rewriting as 
\be
\vev{B^n}_\alpha'=(e^{i\alpha}-1)\,\frac{1}{\sqrt{2\pi}}\int d\phi\,
\left[\left(i\frac{\der}{\der\Phi}\right)^n\,e^{-\frac12\Phi^2}\right]_{\Phi=W'(\phi)},
\ee
we can express it by the integral of the Hermitian polynomials: 
\be
\vev{B^n}_\alpha'=(e^{i\alpha}-1)\,\frac{(-i)^n}{\sqrt{2\pi}}\int^\infty_{-\infty} d\phi\,H_n(W'(\phi))\,
e^{-\frac12 W'(\phi)^2}, 
\label{Bn_unVEV_hermitian}
\ee
where the Hermitian polynomials are defined by 
\be
H_n(x) \equiv (-1)^n e^{\frac12 x^2}\frac{d^n}{dx^n}\,e^{-\frac12 x^2}. 
\ee

The expectation value of $B^n$ normalized by $Z_\alpha$: 
\be
\vev{B^n}_\alpha\equiv \frac{1}{Z_\alpha}\,\vev{B^n}_\alpha'
\ee
trivially vanishes as turning off $\alpha$ in the case that the degree $p$ of $W'(\phi)$ (\ref{W'}) is odd, 
in which the SUSY is not spontaneously broken ($Z_0\neq 0$). 
However, for even $p$ where the SUSY is broken ($Z_0=0$), 
taking the ratio of (\ref{Z_alpha}) and (\ref{Bn_unVEV_hermitian}) we obtain 
\be
\vev{B^n}_\alpha = \frac{(-i)^n\int^\infty_{-\infty} d\phi\,H_n(W'(\phi))\,e^{-\frac12 W'(\phi)^2}}{\int^\infty_{-\infty} d\phi\,e^{-\frac12 W'(\phi)^2}}, 
\ee
which can take a nontrivial value. 
Note that, because the factors $(e^{i\alpha}-1)$ appearing in the numerator and the denominator cancel each other, 
the value of $\vev{B^n}_\alpha$ is not dependent on $\alpha$~\cite{Kuroki:2009yg}.  

The argument so far presented for $T=1$ can be extended to the case of general $T$. 
We put the discussion in appendix~\ref{app:loc_generalT}.

\section{Change of variables and localization in SUSY matrix models}
\label{sec:loc_in_MQM}
\setcounter{equation}{0}
In this section, we discuss localization in  
SUSY matrix models, which yields some new features not seen in the previous section. 
Let us begin with a matrix-model analog of \eqref{dQM_S2}
\bea
S 
& = & Q \sum_{t=1}^TN\tr \,\bar{\psi}(t) \left\{ \frac{i}{2} B(t) 
 -\left(\phi(t+1) -\phi(t) +W'(\phi(t))\right)\right\} 
\nonumber \\
 & = & \sum_{t=1}^TN\tr\, 
\left[\frac12 B(t)^2 +iB(t)\left\{\phi(t+1)-\phi(t)+W'(\phi(t))\right\} \right. \nn \\
 &  &\hspace{3cm} 
\left. \frac{}{} +\bar{\psi}(t) \left\{\psi(t+1) -\psi(t) +QW'(\phi(t))\right\} \right],  
 \label{MQM_S2}
\eea
where all variables are $N\times N$ Hermitian matrices~\footnote{
{}From the viewpoint of spontaneous SUSY breaking in discretized noncritical superstrings, 
SUSY matrix quantum mechanics with a cubic superpotential and an $\cN=1/2$ version of 
the $T=1$ case of (\ref{MQM_S2}) are discussed in  \cite{Marinari:1990jc} and \cite{Nojiri:1992zu}, 
respectively. 
}. Under the periodic boundary condition, 
this action is manifestly 
invariant under $Q$ transformation defined in \eqref{QSUSY}. 

We will focus on the simplest case $T=1$ below. Under the twisted boundary condition \eqref{TBC_alpha}, 
the action is  
\bea
S_\alpha & = & N \tr \left[\frac12 B^2 +iBW'(\phi) 
+\bar{\psi}\left(e^{i\alpha}-1\right)\psi +\bar\psi QW'(\phi)
\right], 
\label{S_TBC}
\eea
and the partition function is defined by
\be
Z_\alpha  \equiv    \left(-1\right)^{N^2}
\int d^{N^2}B \,d^{N^2}\phi \,\left(d^{N^2}\psi \,d^{N^2}\bar{\psi}\right)\, e^{-S_\alpha},  
\label{MMZ_TBC}
\ee
where we fix the normalization of the measure as 
\be
\int d^{N^2}\phi \, e^{-N\tr \,(\frac12 \phi^2)} = \int d^{N^2}B \, e^{-N\tr \,(\frac12 B^2)} = 1, \qquad 
(-1)^{N^2} \int \left(d^{N^2}\psi \,d^{N^2}\bar{\psi}\right)\, e^{-N\tr \,(\bar{\psi}\psi)}=1. 
\label{normalization}
\ee
Explicitly, when $W'(\phi)$ is given as in \eqref{W'}, \eqref{S_TBC} becomes 
\begin{align}
S_\alpha = N \tr \left[\frac12 B^2 +iBW'(\phi) 
+\bar{\psi}\left(e^{i\alpha}-1\right)\psi 
+\sum_{k=1}^pg_k\sum_{\ell=0}^{k-1} \bar{\psi}\,\phi^\ell\, \psi \,\phi^{k-\ell-1}
\right].
\end{align}
Notice the ordering of the matrices in the last term. 
We see that the effect of the external field again remains even after the reduction 
to zero dimension. When $\alpha=0$, $S_{\alpha=0}$ is invariant 
under $Q$ and $\bar{Q}$ given in \eqref{QSUSY0dim} and \eqref{Qbar_SUSY0dim},
both of which become broken explicitly in $S_\alpha$ by introducing the external field $\alpha$.

Now let us discuss localization of the integration in $Z_\alpha$. 
Some aspects are analogous to the discretized SUSY quantum mechanics 
with $T\geq2$ under the identification $N^2=T$ from the viewpoint of systems 
possessing multi-degrees of freedom, 
while there are also interesting new phenomena specific to matrix models.    
We make a change of variables 
\begin{align}
\phi=\tilde\phi+\bar\epsilon\psi,~~~
\bar\psi=\tilde{\bar\psi} -i\bar\epsilon B ,
\label{newvariables}
\end{align}
where in the second equation, $\tilde{\bar\psi}$ satisfies 
\begin{align}
N\tr(B\tilde{\bar\psi})=0,
\label{constraint_MM}
\end{align}
namely, $\tilde{\bar\psi}$ is orthogonal to $B$ with respect to the inner product 
$(A_1, A_2)\equiv N\tr(A_1^\dagger A_2)$. Let us take a basis of $N\times N$ Hermitian matrices 
$\{t^a\}$ ($a=1,\cdots ,N^2$) to be 
orthonormal with respect to the inner product: $N\tr(t^at^b)=\delta_{ab}$. 
More explicitly, we take
\begin{align}
\bar\epsilon\equiv i\frac{\tr(B\bar\psi)}{\tr B^2} = \frac{i}{\cN_B^{\, 2}}\,N\tr(B\bar\psi)
\end{align}
with $\cN_B\equiv ||B|| = \sqrt{N\tr(B^2)}$ the norm of the matrix $B$. 
Notice that in the present case ($N$ is general) $\bar\psi$ is an $N\times N$ matrix and that  
$\bar\epsilon$ does not have enough degrees of freedom to parametrize the whole space 
of $\bar\psi$, 
which is in contrast with the $N=1$ case \eqref{fieldredef} but analogous to the discretized quantum mechanics 
with $T\geq 2$ (\ref{fieldredefT}). 
In fact, $\bar\epsilon$ is used to parametrize a single component of $\bar\psi$ parallel to $B$. 

If we write (\ref{MMZ_TBC}) as 
\be
Z_\alpha = \int d^{N^2}B\,\Xi_\alpha(B), \qquad \Xi_\alpha(B)\equiv \left(-1\right)^{N^2}
\int d^{N^2}\phi \,\left(d^{N^2}\psi \,d^{N^2}\bar{\psi}\right)\, e^{-S_\alpha}, 
\label{Z_Xi_MM}
\ee
and consider the change of the variables in $\Xi_\alpha(B)$, $B$ may be regarded as an external 
variable.  
The measure $d^{N^2}\bar\psi$ can be expressed by the measures associated with 
$\tilde{\bar\psi}$ and $\bar\epsilon$ as 
\begin{align}
d^{N^2}\bar\psi=\frac{i}{\cN_B}\,d\bar\epsilon\,d^{N^2-1}\tilde{\bar\psi},
\label{newmeasure}
\end{align} 
where $d^{N^2-1}\tilde{\bar\psi}$ is explicitly given by introducing 
the constraint (\ref{constraint_MM}) as a delta-function: 
\bea
d^{N^2-1}\tilde{\bar\psi} & \equiv & (-1)^{N^2-1} d^{N^2}\tilde{\bar\psi}\,\delta\left(\frac{1}{\cN_B}\,N\tr(B\tilde{\bar\psi})\right) \nn \\
  & = & (-1)^{N^2-1} \left(\prod_{a=1}^{N^2} d\tilde{\bar\psi}^a \right)\,\frac{1}{\cN_B}\sum_{a=1}^{N^2}B^a \tilde{\bar\psi}^a.
\label{d_psi_tilde_bar}  
\eea
$\tilde{\bar\psi}^a$ and $B^a$ are coefficients 
in the expansion of $\tilde{\bar{\psi}}$ and $B$ by the basis $\{t^a\}$: 
\be
\tilde{\psi}=\sum_{a=1}^{N^2} \tilde{\bar\psi}^a t^a, \qquad B= \sum_{a=1}^{N^2}B^a t^a. 
\ee  
((\ref{newmeasure}) and (\ref{d_psi_tilde_bar}) are analogous to (\ref{d_psi_barT}) and (\ref{measure_psibarT}) 
in the discretized quantum mechanics with $T\geq 2$, respectively.)  
Notice that the measure 
on the RHS of (\ref{newmeasure}) depends on $B$. 
When $B\neq \bm 0$, we can safely change the variables as in \eqref{newvariables} 
and in terms of them the action becomes  
\begin{align}
S_\alpha=N\tr\left[\frac12B^2+iBW'(\tilde\phi)+\tilde{\bar\psi}\left((e^{i\alpha}-1)\psi
+QW'(\tilde\phi)\right)
-(e^{i\alpha}-1)i\bar\epsilon B\psi\right]
\label{newaction1}
\end{align}
with $Q\tilde\phi=\psi$. 

\subsection{$\alpha=0$ case}
Let us first consider the case of the periodic boundary condition ($\alpha=0$).  
Similarly to \eqref{S0_fieldredefT}, 
$S_{\alpha=0}$ does not depend on $\bar\epsilon$ as a consequence of its SUSY invariance, 
because \eqref{newvariables} reads      
\begin{align}
\phi=\tilde\phi+\epsilon Q\tilde\phi,~~~
\bar\psi=\tilde{\bar\psi}+\bar\epsilon Q\tilde{\bar\psi}.
\end{align}
Therefore, 
the contribution to the partition function from $B\neq \bm 0$ 
\be
\tilde{Z}_{\alpha=0} = \int_{||B||\geq \varepsilon} d^{N^2}B\,\Xi_{\alpha=0}(B) \qquad (0<\varepsilon \ll 1)
\ee 
vanishes due to the integration over $\bar\epsilon$ 
according to \eqref{newmeasure}. 
Namely, when $\alpha=0$, the path integral of the partition function \eqref{MMZ_TBC} 
is localized to $B=\bm 0$. 

For the contribution to the partition function from the vicinity of $B=\bm 0$ 
\be
Z_{\alpha=0}^{(0)} = \int_{||B||< \varepsilon} d^{N^2}B\,\Xi_{\alpha=0}(B), 
\label{Z^0_0_MM}
\ee 
we can repeat the same argument as in the previous section. 
For instance, when $W'(\phi)$ is given by (\ref{W'}) of degree $p\geq 2$, rescaling as
\be
\tilde{\phi} = \cN_B^{-\frac{1}{p}}\phi', \qquad \tilde{\bar\psi}=\cN_B^{\frac{p-1}{p}}\bar{\psi}', 
\label{rescale_MM}
\ee
we obtain 
\bea
Z_{\alpha=0}^{(0)} & = & i\left(\frac{-1}{\sqrt{2\pi}}\right)^{N^2} \left(\int^\varepsilon_0d\cN_B\,
\frac{1}{\cN_B^{1+\frac{1}{p}}}\,e^{-\frac12 \cN_B^2}\right) \,
\int d\Omega_B\int d^{N^2}\phi' \,e^{-iN\tr \left(\Omega_B g_p\phi'^p\right)} \nn \\
& & \hspace{7mm} \times  \int d^{N^2}\psi\int d\bar\epsilon \,d^{N^2-1}\bar\psi'\, 
e^{-N\tr\left[\bar{\psi}'g_p\sum_{\ell=0}^{p-1} \phi'^\ell\psi\phi'^{p-\ell-1}\right]} \,\left[1+\cO(\varepsilon^{1/p})\right], 
\eea
where the measure of the $B$-integral was expressed in terms of polar coordinates in ${\bf R}^{N^2}$ as 
\be
d^{N^2}B = \prod_{a=1}^{N^2}\frac{dB^a}{\sqrt{2\pi}} 
= \left(\frac{1}{2\pi}\right)^{\frac{N^2}{2}}\,\cN_B^{N^2-1} d\cN_B \,d\Omega_B, 
\ee 
and $\Omega_B\equiv \frac{1}{\cN_B}\,B$ represents a unit vector in ${\bf R}^{N^2}$. 
Since the $\bar\epsilon$-integral vanishes while the integration of $\cN_B$ becomes singular at the origin,  
$Z_{\alpha=0}^{(0)}$ takes an indefinite form ($\infty\times 0$). 
When $W'(\phi)$ is linear ($p=1$), the $\tilde\phi$-integrals in (\ref{Z^0_0_MM}) yield 
\bea
Z_{\alpha=0}^{(0)} & = & i\left(\frac{-1}{|g_1|}\right)^{N^2} \,
\int_{||B||<\varepsilon} \left(\prod_{a=1}^{N^2}dB^a\right)\,\frac{1}{\cN_B}\,e^{-\frac12 \cN_B^2} \, 
\prod_{a=1}^{N^2}\delta(B^a) \nn \\
& & \times \int d^{N^2}\psi\,\int d\bar\epsilon \,d^{N^2-1}\tilde{\bar\psi} \,
e^{-N\tr (\tilde{\bar\psi}g_1\psi)}, 
\label{Z^0_0_MM_linear}
\eea
which is also of indefinite form -- the $B$-integrals diverge 
while $\int d\bar\epsilon$ trivially vanishes. 
Thus the change of variables (\ref{newvariables}) is not suitable to evaluate $Z^{(0)}_{\alpha=0}$ which 
possibly takes a nonzero value.  

\subsubsection{Unnormalized expectation values} 
Next, let us consider the unnormalized expectation values of $\frac{1}{N}\tr B^n$ ($n\geq 1$): 
\be
\vev{\frac{1}{N}\tr B^n}' \equiv \int d^{N^2}B\,\left(\frac{1}{N}\tr B^n\right) \,\Xi_{\alpha=0}(B).
\ee
Since contribution from the region $||B||\geq \varepsilon$ 
is shown to be zero by the change of variables (\ref{newvariables}), 
we focus on the $B$-integration around the origin ($||B||<\varepsilon$). 

When $W'(\phi)$ is a polynomial (\ref{W'}) of degree $p\geq 2$, after the rescaling (\ref{rescale_MM}) 
we obtain 
\bea
\vev{\frac{1}{N}\tr B^n}' & = & i\left(\int^\varepsilon_0 d\cN_B\,\cN_B^{\,n-1-\frac{1}{p}} \,e^{-\frac12 \cN_B^2} 
\right)\,Y_N \,\left[1+ \cO(\varepsilon^{1/p})\right], \nn \\
Y_N & \equiv & \left(\frac{-1}{\sqrt{2\pi}}\right)^{N^2} \int d\Omega_B \,\frac{1}{N}\tr\left(\Omega_B^n\right) 
\,\int d^{N^2}\phi' \,e^{-iN\tr (\Omega_B g_p \phi'^p)} \nn \\
 & & \times \int d^{N^2}\psi \,\int d\bar\epsilon \,d^{N^2-1}\bar\psi'\, 
e^{-N\tr\left[\bar\psi' g_p \sum_{\ell=0}^{p-1}\phi'^\ell \psi \phi'^{p-\ell-1}\right]}.    
\label{Y_N}
\eea
The $\cN_B$-integral is finite, and it is shown that $Y_N$ definitely vanishes in appendix~\ref{app:Y_N}.   
Thus, the change of variables (\ref{newvariables}) is possible for any $B$ in evaluating 
$\vev{\frac{1}{N}\tr B^n}'$ to give the result 
\be
 \vev{\frac{1}{N}\tr B^n}'=0 \qquad (n\geq 1). 
\label{unvev_Bn_MM} 
\ee
When $W'(\phi)$ is linear, $\vev{\frac{1}{N}\tr B^n}'$ has the same expression 
as the RHS of (\ref{Z^0_0_MM_linear}) except the integrand multiplied by $\frac{1}{N}\tr B^n$. 
It leads to a finite result of the $B$-integration for $n\geq 1$, and (\ref{unvev_Bn_MM}) is also obtained.      
 
Furthermore, it can be similarly shown that the unnormalized expectation values 
of multi-trace operators $\prod_{i=1}^k\frac{1}{N}\tr\,B^{n_i}$ ($n_1, \cdots, n_k\geq 1$) vanish: 
\be
\vev{\prod_{i=1}^k\frac{1}{N}\tr\,B^{n_i}}'=0.
\label{unvev_Bnk_MM}
\ee

\subsubsection{Localization to $W'(\phi)=0$, and localization versus Vandermonde}
Since (\ref{unvev_Bnk_MM}) means 
\be
\vev{e^{-N\tr\left(\frac{u-1}{2}\,B^2\right)}}' 
= \sum_{n=0}^\infty \frac{1}{n!}\,\left(-N^2\frac{u-1}{2}\right)^n \vev{\left(\frac{1}{N}\tr\,B^2\right)^n}'
= \vev{1}'=Z_{\alpha=0}
\ee
for an arbitrary parameter $u$, 
we may compute $\vev{e^{-N\tr\left(\frac{u-1}{2}\,B^2\right)}}'$ to evaluate the partition function 
$Z_{\alpha=0}$.  
It is independent of the value of $u$, so $u$ can be chosen to a convenient value to make the evaluation easier.   
 
Taking $u>0$ and integrating $B$ first, we obtain 
\be
Z_{\alpha=0} =  (-1)^{N^2}\int d^{N^2}\phi\,\left(\frac{1}{u}\right)^{\frac{N^2}{2}}\,
e^{-N\tr \left[\frac{1}{2u}\,W'(\phi)^2\right]} 
\, \int d^{N^2}\psi\,d^{N^2}\bar\psi\, e^{-N\tr\left[\bar\psi QW'(\phi)\right]}. 
\label{Z_0_MM_loc}
\ee 
Then, let us consider the $u\to 0$ limit. 
Localization to $W'(\phi)=0$ takes place because 
\be
\lim_{u\to 0}\left(\frac{1}{u}\right)^{\frac{N^2}{2}}\,e^{-N\tr \left[\frac{1}{2u}\,W'(\phi)^2\right]}
= (2\pi)^{\frac{N^2}{2}}\,\prod_{a=1}^{N^2}\delta (W'(\phi)^a). 
\ee
It is important to recognize that $W'(\phi)^a=0$ for all $a$ implies localization 
to a continuous space. Namely, if this condition is met, $W'(U^\dagger \phi U)^a=0$ 
for ${}^\forall U\in SU(N)$.  Thus the original $SU(N)$ gauge symmetry in the matrix model 
makes the localization continuous in nature. This is characteristic  
of SUSY matrix models.

The observation above suggests that in order to localize 
the path integral to discrete points, we should switch to a description 
in terms of gauge invariant quantities.   This motivates us 
to change the expression of $\phi$ to its eigenvalues and $SU(N)$ angles as 
\be
\phi=U \begin{pmatrix} \lambda_1 &     &     \\
                                & \ddots &     \\
                                 &       &  \lambda_N \end{pmatrix} U^\dagger, \qquad U\in SU(N).  
\label{phi_lambda_U}                                 
\ee 
This leads to an interesting situation, which is peculiar to SUSY matrix models and is
not seen in the discretized SUSY quantum mechanics.    
For a polynomial $W'(\phi)$ given by (\ref{W'}),  
the partition function (\ref{Z_0_MM_loc}) becomes 
\be
Z_{\alpha=0}=\left(\frac{1}{u}\right)^{\frac{N^2}{2}}\,\int d^{N^2}\phi\,e^{-N\tr \left[\frac{1}{2u}\,W'(\phi)^2\right]} 
\det\left[\sum_{k=1}^pg_k\sum_{\ell=0}^{k-1}\phi^\ell\otimes \phi^{k-\ell-1}\right],
\label{Z_0_MM_loc2}
\ee
after the Grassmann integrals. Note that the $N^2\times N^2$ matrix 
$\sum_{k=1}^pg_k\sum_{\ell=0}^{k-1}\phi^\ell\otimes \phi^{k-\ell-1}$ has the eigenvalues 
$\sum_{k=1}^pg_k\sum_{\ell=0}^{k-1}\lambda_i^\ell\lambda_j^{k-\ell-1}$ ($i,j =1, \cdots, N$). 
Thus, the fermion determinant can be expressed as 
\bea
\det\left[\sum_{k=1}^pg_k\sum_{\ell=0}^{k-1}\phi^\ell\otimes \phi^{k-\ell-1}\right]
& = & \prod_{i,j=1}^N \left[\sum_{k=1}^pg_k\sum_{\ell=0}^{k-1}\lambda_i^\ell\lambda_j^{k-\ell-1}\right] \nn \\
& = & \left(\prod_{i=1}^N W''(\lambda_i)\right)\,
\prod_{i>j} \left(\frac{W'(\lambda_i)-W'(\lambda_j)}{\lambda_i-\lambda_j}\right)^2. 
\eea
The measure $d^{N^2}\phi$ given in \eqref{normalization} can be also recast to
\be
 d^{N^2}\phi = \tilde{C}_N \Bigl(\prod_{i=1}^N d\lambda_i \Bigr)\,\triangle(\lambda)^2\, dU, 
\ee
where $\triangle(\lambda)=\prod_{i>j}(\lambda_i-\lambda_j)$ is the Vandermonde determinant, and 
$dU$ is the $SU(N)$ Haar measure normalized by $\int dU=1$. 
$\tilde{C}_N$ is a numerical factor depending only on $N$ determined by 
\be
\frac{1}{\tilde{C}_N} = \int \Bigl(\prod_{i=1}^N d\lambda_i\Bigr) \,\triangle(\lambda)^2\, 
e^{-N\sum_{i=1}^N \frac12 \lambda_i^2}.
\label{CtildeN}
\ee  
Plugging these into (\ref{Z_0_MM_loc2}), we obtain 
\bea
Z_{\alpha=0} & = & \tilde{C}_N\,\int \Bigl(\prod_{i=1}^Nd\lambda_i\Bigr)\, 
 \left(\prod_{i=1}^N W''(\lambda_i)\right)\,
\left\{\prod_{i>j} \frac{1}{u}\,\left(W'(\lambda_i)-W'(\lambda_j)\right)^2\right\} \nn \\
& & \hspace{28mm} \times 
\left(\frac{1}{u}\right)^{\frac{N}{2}}\,e^{-N\sum_{i=1}^N\frac{1}{2u}\,W'(\lambda_i)^2}. 
\label{Z_0_MM_loc3}
\eea
In this expression, the factor in the second line  
forces eigenvalues to be localized at the critical points of the superpotential as $u\to 0$, while 
the last factor in the first line, 
which is proportional to the square of the Vandermonde determinant of $W'(\lambda_i)$, 
gives repulsive force among eigenvalues which prevents them from collapsing to the critical points.   
The dynamics of eigenvalues is thus determined by balance of the attractive force 
to the critical points originating from the localization and the repulsive force 
from the Vandermonde determinant.   
This kind of dynamics has not been seen in the discretized SUSY quantum mechanics 
discussed in the previous section.  

To proceed with the analysis, 
let us consider the situation of each eigenvalue $\lambda_i$ fluctuating around the critical point $\phi_{c,i}$:
\be
\lambda_i = \phi_{c, i}+\sqrt{u}\,\tilde{\lambda}_i \qquad (i=1, \cdots, N),
\ee
where $\tilde{\lambda}_i$ is a fluctuation, and $\phi_{c,1}, \cdots, \phi_{c, N}$ are allowed 
to coincide with each other. 
Then, the partition function (\ref{Z_0_MM_loc3}) takes the form 
\bea
Z_{\alpha=0} & = & \tilde{C}_N \sum_{\phi_{c,i}} \,\int \Bigl(\prod_{i=1}^Nd\tilde{\lambda_i}\Bigr)\,
\prod_{i=1}^NW''(\phi_{c, i})\, 
\prod_{i>j}\left(W''(\phi_{c,i})\tilde{\lambda}_i-W''(\phi_{c,j})\tilde{\lambda}_j\right)^2 \nn \\
& & \hspace{28mm} \times e^{-N\sum_{i=1}^N \frac12 W''(\phi_{c, i})^2\tilde{\lambda}_i^2} 
+\cO(\sqrt{u}). 
\label{Z_0_MM_loc4}
\eea
Although only the Gaussian factors become relevant as $u\to 0$ similarly to 
the $N=1$ case (\ref{Z_0_T=1_loc_3}),  
there remain $N(N-1)$-point vertices originating from the Vandermonde determinant 
of $W'(\lambda_i)$ which yield a specific effect of SUSY matrix models.  
Before computing (\ref{Z_0_MM_loc4}) for a general case, 
let us consider the following two simple cases. 

\paragraph{Gaussian case}
In the case of $W'(\phi)=g_1\phi$, where the corresponding  scalar potential 
$\frac12 W'(\phi)^2$ is Gaussian, 
the critical point is only the origin: $\phi_{c,1}=\cdots=\phi_{c,N}=0$. 
Then, (\ref{Z_0_MM_loc4}) is reduced to 
\be
Z_{\alpha=0} = \tilde{C}_N \int \Bigl(\prod_{i=1}^N d\tilde{\lambda}_i\Bigr)\,g_1^{N^2} \,
\prod_{i>j} \left(\tilde{\lambda}_i-\tilde{\lambda}_j\right)^2\, e^{-N\sum_{i=1}^N\frac12 g_1^2\tilde{\lambda}_i^2}, 
\ee  
where no $\cO(\sqrt{u})$ term appears since $W'(\phi)$ is linear. 
By using (\ref{CtildeN}) we obtain the result nothing but eq. (B.3) in~\cite{Kuroki:2009yg}: 
\be
Z_{\alpha=0}=({\rm sgn}(g_1))^{N^2} = ({\rm sgn}(g_1))^{N}.  
\ee

\paragraph{Double-well case}
For $W'(\phi)=g(\phi^2-\mu^2)$ with $\mu>0$, which gives a scalar potential 
of double-well shape, each of $\phi_{c,i}$ is equal to $\mu$ or $-\mu$.  
Let us consider the case that the first $\nu_+N$ eigenvalues are around $\mu$ 
and the remaining $\nu_-N$ around $-\mu$:
\be
\phi_{c,1}=\cdots =\phi_{c,\nu_+N}=\mu, \qquad 
\phi_{c, \nu_+N+1}=\cdots = \phi_{c, N}=-\mu,
\label{nu+_nu-}
\ee
where the filling fractions $\nu_+, \nu_-$ satisfy $\nu_+ +\nu_- =1$. 
Let $Z_{(\nu_+, \nu_-)}$ be a contribution to the partition function $Z_{\alpha=0}$ 
from small fluctuations around (\ref{nu+_nu-}). Then, 
\be
Z_{\alpha=0} = \sum_{\nu_+N=0}^N \frac{N!}{(\nu_+N)!(\nu_-N)!}\, Z_{(\nu+, \nu_-)} . 
\label{Z_0_MM_total}
\ee 
Since
\bea
& & W''(\phi_{c,i}) =2g\mu \qquad (i=1, \cdots, \nu_+N), \nn \\
& & W''(\phi_{c,i}) =-2g\mu \qquad (i=\nu_+N+1, \cdots, N), 
\eea
we have 
\bea
Z_{(\nu_+, \nu_-)} & = &  (-1)^{\nu_-N}(2g\mu)^{N^2}\,\tilde{C}_N 
\int \Bigl(\prod_{i=1}^N d\tilde{\lambda}_i\Bigr)\,e^{-N\sum_{i=1}^N\frac12 (2g\mu)^2\tilde{\lambda}_i^2} \nn \\
 & & \times \prod_{\nu_+N\geq i>j\geq 1} \left(\tilde{\lambda}_i-\tilde{\lambda}_j\right)^2 
 \prod_{N\geq i>j\geq\nu_+N+1}\left(\tilde{\lambda}_i-\tilde{\lambda}_j\right)^2 \nn \\
 & & \times \prod_{N\geq i\geq \nu_+N+1,\, \nu_+N\geq j\geq 1}\left(\tilde{\lambda}_i+\tilde{\lambda}_j\right)^2
+\cO(\sqrt{u}).
\eea 
Note that flipping the sign  
$\tilde{\lambda}_i \to -\tilde{\lambda}_i$ for $i=\nu_+N+1, \cdots, N$ 
makes the factors in the second and third lines combined to the square of the single Vandermonde determinant 
$\triangle(\tilde{\lambda})^2$. Thus, 
$Z_{(\nu_+, \nu_-)}$ can be expressed by the partition function of the Gaussian SUSY matrix model 
with $g_1=2g\mu$ : 
\bea
Z_{(\nu_+, \nu_-)} & = &  (-1)^{\nu_-N}(2g\mu)^{N^2}\,\tilde{C}_N 
\int \Bigl(\prod_{i=1}^N d\tilde{\lambda}_i\Bigr)\,\triangle(\tilde{\lambda})^2\,
e^{-N\sum_{i=1}^N\frac12 (2g\mu)^2\tilde{\lambda}_i^2} +\cO(\sqrt{u})\nn \\
& = & (-1)^{\nu_-N}\,({\rm sgn}(2g\mu))^N +\cO(\sqrt{u}). 
\label{Z_nu+_nu-}
\eea 

Here, let $Z_{G, \nu_\pm}$ be the partition functions of the Gaussian SUSY matrix models with the matrix size 
$\nu_\pm N\times \nu_\pm N$ describing contributions from Gaussian fluctuations around the minima $\phi= \pm\mu$, 
respectively. 
Since 
\be
Z_{G, \nu_+}=({\rm sgn}(2g\mu))^{\nu_+N}, \qquad Z_{G,\nu_-}=({\rm sgn}(-2g\mu))^{\nu_-N}, 
\ee
we can show 
\be
Z_{(\nu_+,\nu-)}= Z_{G, \nu_+}\,Z_{G,\nu_-}
\label{factorization}
\ee
in the limit $u\to 0$. It holds for arbitrary $N$, and leads to the statement (4.17) in the previous 
paper~\cite{Kuroki:2009yg} in the large-$N$ limit. 
Note that the integrand in the first line of (\ref{Z_nu+_nu-}) cannot be factorized into 
the products of two functions -- 
one is a function of $\tilde{\lambda}_i$ ($i=1, \cdots, \nu_+N$) 
and the other of $\tilde{\lambda}_i$ ($i=\nu_+N+1, \cdots, N$) -- due to the Vandermonde determinant. 
Nevertheless, the factorization (\ref{factorization}) takes place at the level of the partition function.
It is interesting to get more insight about the factorization, which will be useful to make deeper 
our understanding on the 
structure of SUSY matrix models.  
 
Finally, we find that the total partition function (\ref{Z_0_MM_total}) vanishes:
\be
Z_{\alpha=0} =  \sum_{\nu_-N=0}^N \frac{N!}{(\nu_+N)!(\nu_-N)!}\, Z_{(\nu_+,\nu_-)} 
=  ({\rm sgn}(2g\mu))^N\, \left(1+(-1)\right)^N = 0, 
\ee
which is expected from the spontaneous SUSY breaking in the case of double-well scalar potentials 
at finite $N$.

\paragraph{General case}
Now, let us evaluate (\ref{Z_0_MM_loc4}) for a general superpotential. 
We change the integration variables as 
\be
\tilde{\lambda}_i=\frac{1}{W''(\phi_{c,i})}\,y_i, 
\label{lambda_y}
\ee
then the integration of $\tilde{\lambda}_i$ becomes 
$\int^\infty_{-\infty}d\tilde{\lambda}_i\cdots = \frac{1}{|W''(\phi_{c,i})|}\int^\infty_{-\infty}dy_i\cdots$. 
In the limit $u\to 0$, (\ref{Z_0_MM_loc4}) is computed to be 
\bea
Z_{\alpha=0} & = & \sum_{\phi_{c,i}}\prod_{i=1}^N\frac{W''(\phi_{c,i})}{|W''(\phi_{c,i})|}\,
\left\{\tilde{C}_N\int^\infty_{-\infty}\Bigl(\prod_{i=1}^Ndy_i\Bigr)\,\triangle(y)^2\,
e^{-N\sum_{i=1}^N\frac12 y_i^2}
\right\}
\nn \\
& = & \sum_{\phi_{c,i}}\prod_{i=1}^N{\rm sgn}\left(W''(\phi_{c,i})\right) \nn \\
& = & \left[\sum_{\phi_c:\,W'(\phi_c)=0}{\rm sgn}\left(W''(\phi_{c})\right)\right]^N. 
\label{Z_0_MM_loc5}
\eea
Note that the last factor in the first line of (\ref{Z_0_MM_loc5}) is nothing but the partition function 
of the Gaussian case with $g_1=1$. 
The last line of (\ref{Z_0_MM_loc5}) tells that the total partition function is given by the $N$-th power of 
the $N=1$ case (\ref{Z_0_T=1_loc_2})~\footnote{ 
Since $\sum_{\phi_c:\,W'(\phi_c)=0}{\rm sgn}\left(W''(\phi_{c})\right)$ is equal to 
the mapping degree (\ref{mapping_degree}) from (\ref{mapping_degree_localization}), 
$Z_{\alpha=0}$ is also expressed as the $N^2$-th power of the $N=1$ case. 
It is analogous to (\ref{localization_T}) in the discretized SUSY quantum mechanics 
with the identification $N^2=T$. 
}. 

Furthermore, we consider a case that the superpotential $W(\phi)$ has $K$ 
nondegenerate critical points $a_1, \cdots, a_K$. 
Namely, $W'(a_I)=0$ and $W''(a_I)\neq 0$ for each $I=1, \cdots, K$. 
The scalar potential $\frac12 W'(\phi)^2$ has $K$ minima at $\phi=a_1, \cdots, a_K$. 
When $N$ eigenvalues are fluctuating around the minima, 
we focus on the situation that \\
\noindent
$\lambda_i$ ($i=1, \cdots, \nu_1N$) are around $\phi=a_1$\\
$\lambda_{\nu_1N+i}$ ( $i=1, \cdots, \nu_2N$) are around $\phi=a_2$\\
\hspace{28mm}$\cdots$\\
$\lambda_{\nu_1N+\cdots+\nu_{K-1}N+i}$ ($i=1, \cdots, \nu_KN$) are around $\phi=a_K$,\\
where $\nu_1, \cdots, \nu_K$ are filling fractions satisfying $\sum_{I=1}^K\nu_I=1$.   
Let $Z_{(\nu_1, \cdots, \nu_K)}$ be a contribution to the total partition function $Z_{\alpha=0}$ 
from the above configuration. Then, 
\be
Z_{\alpha=0} = \sum_{\nu_1N, \cdots, \nu_KN=0}^N\frac{N!}{(\nu_1N)!\cdots (\nu_KN)!}\,
Z_{(\nu_1, \cdots, \nu_K)}. 
\ee
(The sum is taken under the constraint $\sum_{I=1}^K\nu_I=1$.)
Since $Z_{(\nu_1, \cdots, \nu_K)}$ is equal to the second line of (\ref{Z_0_MM_loc5}) with 
$\phi_{c,i}$ fixed as 
\bea
& & \phi_{c,1}=\cdots =\phi_{c,\nu_1N}=a_1, \nn \\
& & \phi_{c, \nu_1N+1}=\cdots =\phi_{c,\nu_1N+\nu_2N}=a_2, \nn \\
& & \cdots \nn \\
& & \phi_{c, \nu_1N+\cdots +\nu_{K-1}N+1} =\cdots = \phi_{c, N} = a_K, 
\eea
we obtain the generalization of the double-well case (\ref{factorization}) : 
\be
Z_{(\nu_1, \cdots, \nu_K)} = \prod_{I=1}^KZ_{G, \nu_I}, \qquad 
Z_{G, \nu_I}= \left({\rm sgn}\left(W''(a_I)\right)\right)^{\nu_IN}. 
\ee
$Z_{G, \nu_I}$ can be interpreted as the partition function of the Gaussian SUSY matrix model with the 
matrix size $\nu_IN\times \nu_IN$ describing contributions from Gaussian fluctuations around 
$\phi=a_I$.

\subsection{$\alpha\neq 0$ case}
 In the presence of the external field $\alpha$, 
let us consider $\Xi_\alpha(B)$ in (\ref{Z_Xi_MM}) with the action (\ref{newaction1}) obtained after 
the change of variables (\ref{newvariables}). 
Using the explicit form of the measure (\ref{newmeasure}) and (\ref{d_psi_tilde_bar}), 
we obtain 
\bea
\Xi_{\alpha}(B) & = & (e^{i\alpha}-1)\,\frac{(-1)^{N^2-1}}{\cN_B^2}\, \int d^{N^2}\tilde\phi\,
\left(d^{N^2}\psi\,d^{N^2}\tilde{\bar\psi}\right)\, 
e^{-N\tr\left[\frac12 B^2 +iBW'(\tilde{\phi})+\tilde{\bar\psi}QW'(\tilde{\phi})\right]} \nn \\
& & \hspace{49mm}\times N\tr(B\tilde{\bar\psi})\,N\tr(B\psi)\,e^{-(e^{i\alpha}-1)\,N\tr(\tilde{\bar\psi}\psi)}, 
\eea  
which is valid for $B\neq \bm 0$. 
Although we can proceed the computation further, 
it is more convenient to invoke another method based on the Nicolai mapping we will present 
in the next section.

\section{$(e^{i\alpha}-1)$-expansion and Nicolai mapping} 
\label{sec:Nicolai}
\setcounter{equation}{0}
In the previous section, we tried to compute the partition function $Z_\alpha$ in the presence of 
the external field. We have seen that the change of variables is useful to localize the path integral, 
but in the $\alpha\neq 0$ case the external field makes the localization incomplete and 
the explicit computation somewhat cumbersome. 
In this section, we instead compute $Z_\alpha$ in an expansion with respect to $(e^{i\alpha}-1)$. 
For the purpose of examining the spontaneous SUSY breaking, we are interested in 
behavior of $Z_\alpha$ in the $\alpha\rightarrow 0$ limit. 
Thus it is expected that it will be often sufficient to compute $Z_\alpha$ in the leading order 
of the $(e^{i\alpha}-1)$-expansion for our purpose.

\subsection{Finite $N$}
\label{subsec:finteN}
Performing the integration over fermions and the auxiliary field $B$ 
in \eqref{MMZ_TBC} with $W'(\phi)$ in (\ref{W'}), we have 
\begin{align}
Z_\alpha=\int d^{N^2}\phi\,
\det\left((e^{i\alpha}-1)\bm 1\otimes\bm 1+\sum_{k=1}^pg_k\sum_{\ell=0}^{k-1}\phi^\ell\otimes\phi^{p-\ell -1}\right)
e^{-N\tr\frac12W'(\phi)^2}. 
\label{Zalpha}
\end{align}
Hereafter, let us expand this with respect to $(e^{i\alpha}-1)$ 
as 
\begin{align}
Z_\alpha=\sum_{k=0}^{N^2}(e^{i\alpha}-1)^k \, Z_{\alpha,k},
\label{Aexpansion}
\end{align}
and derive a formula in the leading order of this expansion. 
The change of variable $\phi$ as (\ref{phi_lambda_U}) recasts (\ref{Zalpha}) to 
\begin{align}
Z_\alpha=\tilde C_N\int\Bigl(\prod_{i=1}^Nd\lambda_i\Bigr)\,\triangle(\lambda)^2\,
\prod_{i,j=1}^N\left(e^{i\alpha}-1+\sum_{k=1}^pg_k\sum_{\ell=0}^{k-1}\lambda_i^\ell\lambda_j^{p-\ell -1}\right)
e^{-N\sum_{i=1}^N\frac12W'(\lambda_i)^2},
\label{Zalphabyeigenvalue}
\end{align}
after the $SU(N)$ angles are integrated out. 
Crucial observation is that we can apply the Nicolai mapping for each $i$ 
even in the presence of the external field 
\begin{align}
\Lambda_i=(e^{i\alpha}-1)\lambda_i+W'(\lambda_i),
\label{Nicolai}
\end{align}
in terms of which the partition function is basically expressed as 
an unnormalized expectation value of the Gaussian matrix model 
\begin{align}
Z_\alpha=\tilde C_N\int \Bigl(\prod_{i=1}^N d\Lambda_i\Bigr)\,\prod_{i>j}(\Lambda_i-\Lambda_j)^2
e^{-N\sum_i\frac12\Lambda_i^2}
e^{-N\sum_i\left(-A\Lambda_i\lambda_i+\frac12A^2\lambda_i^2\right)},
\label{ZbyNicolai}
\end{align}
where $A=e^{i\alpha}-1$. However, there is an important difference from the Gaussian matrix model, which  
originates from the fact that the Nicolai mapping \eqref{Nicolai} is not one to one. 
As a consequence, $\lambda_i$ has several branches as a function of $\Lambda_i$ and 
it has a different expression according to each of the branches. 
Therefore, since the last factor of \eqref{ZbyNicolai} contains $\lambda_i(\Lambda_i)$, 
we have to take account of the branches and divide the integration region of $\Lambda_i$ 
accordingly. Nevertheless, we can derive a rather simple formula at least in the leading order 
of the expansion in terms of $A$ owing to the Nicolai mapping \eqref{Nicolai}. 
In the following, let us concentrate on the cases where 
\be
\Lambda_i\rightarrow \infty \quad \mbox{as}\quad \lambda_i\rightarrow\pm\infty,\qquad \mbox{or} \qquad 
\Lambda_i\rightarrow -\infty \quad \mbox{as}\quad \lambda_i\rightarrow\pm\infty,
\ee
i.e. the leading order of $W'(\phi)$ is even. 
In such cases, 
we can expect spontaneous SUSY breaking, in which the leading nontrivial expansion coefficient is relevant 
since the zeroth order partition function vanishes: $Z_{\alpha=0}=Z_{\alpha, 0}=0$. 
Namely, in the expansion of the last factor in (\ref{ZbyNicolai}) 
\be
e^{-N\sum_{i=1}^N\left(-A\Lambda_i\lambda_i+\frac12A^2\lambda_i^2\right)} 
= 1-N\sum_{i=1}^N\left(-A\Lambda_i\lambda_i+\frac12A^2\lambda_i^2\right) +\cdots, 
\ee
the first term ``1'' does not contribute to $Z_{\alpha}$. It can be understood from the fact 
that it does not depend on the branches and thus the Nicolai mapping 
becomes trivial, i.e. The mapping degree is zero. 
Notice that the second term also gives a vanishing effect.  
For each $i$, we have the unnormalized expectation value of 
$N\left(A\Lambda_i\lambda_i-\frac12A^2\lambda_i^2\right)$, 
where the $\Lambda_j$-integrals  ($j\neq i$) are independent of the branches leading to the trivial 
Nicolai mapping. 
Thus, in order to get a nonvanishing result, 
we need a branch-dependent piece in the integrand for any $\Lambda_i$. 
This immediately shows that in the expansion \eqref{Aexpansion}, 
$Z_{\alpha,k}=0$ for $k=0,.\cdots, N-1$ and that 
the first possibly nonvanishing contribution starts from ${\cal O}(A^N)$ as 
\begin{align}
Z_{\alpha,N}=\left.\tilde C_N \,N^N\int \Bigl(\prod_{i=1}^N d\Lambda_i\Bigr)\,\prod_{i>j}(\Lambda_i-\Lambda_j)^2 \,
e^{-N\sum_{i=1}^N \frac12\Lambda_i^2}\,\prod_{i=1}^N(\Lambda_i\lambda_i)\right|_{A=0}.
\label{coeff}
\end{align}
Note that the $A=(e^{i\alpha}-1)$-dependence of the integrand comes also from $\lambda_i$ 
as a function of $\Lambda_i$ through \eqref{Nicolai}. 
Although the integration over $\Lambda_i$ above should be divided into the branches, 
if we change the integration variables so that we will recover the original $\lambda_i$ 
with $A=0$ (which we call $x_i$) by 
\begin{align}
\Lambda_i=W'(x_i),
\end{align}
then by construction the integration of $x_i$ is standard and runs from $-\infty$ to $\infty$. 
Therefore, we arrive at 
\bea
Z_{\alpha,N}& = & \tilde C_N \,N^N \int^\infty_{-\infty}\Bigl(\prod_{i=1}^Ndx_i\Bigr)\,
\prod_{i=1}^N\left(W^{\prime\prime}(x_i)W'(x_i)x_i\right)\,\prod_{i>j}(W'(x_i)-W'(x_j))^2 \nn \\
& & \hspace{35mm} \times e^{-N\sum_{i=1}^N\frac12W'(x_i)^2},
\label{ZalphaN}
\eea
which does not vanish in general. For example, taking $W'(\phi)=g(\phi^2-\mu^2)$ 
we have for $N=2$
\be
Z_{\alpha,2} = 10g^2\tilde{C}_2 I_0^2 
\left[\frac{I_4}{I_0}-\frac{9}{5}\left(\frac{I_2}{I_0}\right)^2\right],
\ee
where 
\be
I_n \equiv \int^\infty_{-\infty} d\lambda \,\lambda^n \, e^{-g^2(\lambda^2-\mu^2)^2} \qquad (n=0,2,4, \cdots).
\ee
In fact, when $g=1$, $\mu^2=1$ (double-well scalar potential case) 
we find 
\be
I_0 = 1.97373,~~~\frac{I_4}{I_0}-\frac{9}{5}\left(\frac{I_2}{I_0}\right)^2 = -0.165492 \neq 0,
\ee
hence $Z_{\alpha,2}$ actually does not vanish. 
In the case of the discretized SUSY quantum mechanics, we have seen in (\ref{Z_alpha_Z_0_T}) that 
the expansion of $Z_\alpha$ with respect to $(e^{i\alpha}-1)$ 
terminates at the linear order for any $T$. 
Thus, the nontrivial $\cO(A^N)$ contribution of higher order 
can be regarded as a specific feature of SUSY matrix models.  

We stress here that, although we have expanded 
the partition function in terms of $(e^{i\alpha}-1)$ and \eqref{ZalphaN} is 
the leading order one, it is an exact result of the partition function for any finite $N$ 
and any polynomial $W'(\phi)$ of even degree in the presence of the external field. 
Thus, it provides a firm ground for discussion 
of spontaneous SUSY breaking in various settings.

\subsection{Large-$N$}
\label{subsec:largeN}
As an application of \eqref{ZalphaN}, let us discuss SUSY breaking/restoration 
in the large-$N$ limit of our SUSY matrix models. From \eqref{ZalphaN}, 
introducing the eigenvalue density
\begin{align}
\rho(x)=\frac1N\sum_{i=1}^N\delta(x-x_i),
\end{align}
the leading ${\cal O}(A^N)$ part of $Z_\alpha$ is rewritten as 
\be
Z_{\alpha,N}=\tilde C_N\,N^N\int\Bigl(\prod_{i=1}^Ndx_i\Bigr)\,\exp(-N^2F)
\ee
with 
\bea
F & \equiv & - \int dxdy\rho(x)\rho(y)\log\left|W'(x)-W'(y)\right| + \int dx\rho(x)\frac12W'(x)^2 \nn \\
  & & -\frac 1N \int dx\rho(x)\log(W^{\prime\prime}(x)W'(x)x). 
\label{Zalphabyrho}
\eea
In the large-$N$ limit, $\rho(x)$ is given as a solution to the saddle point equation 
obtained from ${\cal O}(1)$ part of $F$ as 
\begin{align}
0=\dashint dy\rho(y)\frac{W^{\prime\prime}(x)}{\left|W'(x)-W'(y)\right|}
-\frac12W'(x)W^{\prime\prime}(x),
\label{SPE}
\end{align}
provided that there exists an ${\cal O}(1)$ solution of this equation. 
Plugging a solution $\rho_0(x)$ into $F$ in \eqref{Zalphabyrho}, we get $Z_\alpha$  
in the large-$N$ limit in the leading order of $(e^{i\alpha}-1)$-expansion as 
\begin{align}
&Z_{\alpha,N}\rightarrow N^N\exp(-N^2F_0), \nonumber \\
&F_0=-\int dxdy\rho_0(x)\rho_0(y)\log\left|W'(x)-W'(y)\right|+\int dx\rho_0(x)\frac12W'(x)^2
-\frac{1}{N^2}\log C_N,
\label{F0}
\end{align}
where $C_N$ is a factor dependent only on $N$ which arises in replacing 
the integration over $\phi$ by the saddle point of its eigenvalue density, 
thus including $\tilde C_N$. From consideration of the Gaussian matrix model, 
$C_N$ is calculated in \cite{Kuroki:2009yg} as 
\begin{align}
C_N=\exp\left[\frac34N^2+{\cal O}(N^0)\right],
\end{align}
and is expected to be independent of the form of superpotential. 
In \eqref{F0} we notice that, if we include 
${\cal O}(1/N)$ part of $F$ (the last term in \eqref{Zalphabyrho}) 
in deriving the saddle point equation, the solution will receive an ${\cal O}(1/N)$ correction 
as $\rho(x)=\rho_0(x)+\frac1N\rho_1(x)$.  
However, when we substitute this into \eqref{Zalphabyrho}, 
$\rho_1(x)$ will contribute to $F$ only by the order ${\cal O}(1/N^2)$, 
because $\cO(1/N)$ corrections to $F_0$ under $\rho_0(x) \to \rho_0(x) + \frac1N \rho_1(x)$ vanish  
as a result of the saddle point equation at the leading order \eqref{SPE} satisfied by $\rho_0(x)$. 

On the other hand, if we set $\alpha=0$ at the level of \eqref{Zalphabyeigenvalue}, 
we have 
\begin{align}
Z_{\alpha=0}=\int\Bigl(\prod_{i=1}^Nd\lambda_i\Bigr)\,\prod_{i=1}^NW''(\lambda_i)\,
\prod_{i>j}(W'(\lambda_i)-W'(\lambda_j))^2 \, e^{-N\sum_{i=1}^N\frac12W'(\lambda_i)^2},
\label{alpha=0}
\end{align}
from which we obtain exactly the same saddle point equation as \eqref{SPE} 
at $N=\infty$. Namely, making the expansion with respect to $(e^{i\alpha}-1)$ 
affects only the subleading part of $F$ in the $1/N$-expansion 
as one can see by comparing \eqref{ZalphaN} and \eqref{alpha=0}. 
It is also the same as the saddle point equation (3.15) in the previous paper \cite{Kuroki:2009yg}. 
Thus, various large-$N$ solutions derived in section 4.1 in \cite{Kuroki:2009yg}, which restore SUSY, 
can be reproduced from 
$Z_{\alpha,N}$ or $Z_\alpha$ in the large-$N$ limit followed by the $\alpha\to 0$ limit, 
in spite of the SUSY breaking at any finite $N$ ($Z_{\alpha=0}=0$). 
Let us see it explicitly for the free energy. 
When $N$ is large but finite, the twisted partition function will take the form 
\be
Z_\alpha \sim (e^{i\alpha}-1)^N N^N \,c\,e^{-N^2F_0-NF_1}, 
\ee
where $\frac{1}{N}\,F_1$ is $\cO(1/N)$ contribution of $F$ given as 
\begin{align}
F_1=-\int dx\rho_0(x)\log(W^{\prime\prime}(x)W'(x)x),
\label{F_1}
\end{align}
and the coefficient $c$ comes from $\cO(1/N^2)$ contribution of $F$. 
The free energy, which corresponds to the quantity $-\frac{1}{N^2}\,\log |Z_\alpha|$ 
leads to $F_0$ in the large-$N$ limit followed by $\alpha\to 0$. 
Notice that, although the effect of the twist $(e^{i\alpha}-1)^N$ is of the subleading order at large $N$, 
it plays a crucial role to obtain the large-$N$ free energy $F_0$.  
(If $\alpha$ was sent to zero before the large-$N$ limit, we would have the vanishing partition function 
and could not find the large-$N$ free energy $F_0$.)


\subsection{Example: SUSY matrix model with double-well potential}
\label{subsec:DWMM}
For illustration of results in the previous subsection, let us consider the SUSY matrix model with 
\begin{align}
W'(\phi)=\phi^2-\mu^2 \qquad (\mu^2 \in {\bf R}).
\label{DW}
\end{align}
In this case the saddle point equation \eqref{SPE} becomes 
\begin{align}
\dashint \frac{\rho(y)}{x-y}+\dashint \frac{\rho(y)}{x+y}=x^3-\mu^2x. 
\label{DWSPE}
\end{align}
In section 4.1 of \cite{Kuroki:2009yg}, we have obtained an asymmetric one-cut solution 
where the eigenvalue density has a single support $[a,b]$ with $b>a>0$ 
and also a two-cut solution with a symmetric support $[-b,-a] \cup [a,b]$~\footnote{
Note that $\mu^2$ here corresponds to $-\mu^2$ in section 4.1 of \cite{Kuroki:2009yg}. 
Interestingly, 
the eigenvalue distribution of the two-cut solution is not ${\bf Z}_2$ symmetric in general. 
In fact, 
\be
\rho_0(x)= \begin{cases} \frac{\nu_+}{\pi}x\sqrt{(x^2-a^2)(b^2-x^2)} \qquad (a<x<b) \\
 \frac{\nu_-}{\pi}|x|\sqrt{(x^2-a^2)(b^2-x^2)} \qquad (-b<x<-a) \end{cases}
\label{two-cut_sol}
\ee
is the explicit form of the solution with the filling fraction ($\nu_+, \nu_-$), which includes the asymmetric 
one-cut solution as a special case $(\nu_+, \nu_-)=(1,0)$.
}. 
Here, $a^2=-2+\mu^2$, $b^2=2+\mu^2$, 
thus they are valid for $\mu^2>2$. 
In $N=1$ case, it is well known that the SUSY is spontaneously broken for \eqref{DW}, 
but in the matrix model case we have shown in \cite{Kuroki:2009yg} that 
the SUSY is restored in the large-$N$ limit for both solutions. In particular, the free energies 
for both solutions are shown to vanish and therefore they coincide with 
the value of the free energy of the Gaussian matrix model. 
It is also proven that the expectation values $\vev{\frac{1}{N}\tr B^n}$ ($n=1, 2, \cdots$) 
are all nil. Here it is worth pointing out 
that the principal value in the saddle point equation, in particular in the second term 
in \eqref{DWSPE} plays a crucial role in the existence of the two-cut solution. 
In this subsection we investigate a one-cut solution with a symmetric support $[-c,c]$ 
which has not been discussed in \cite{Kuroki:2009yg}. 

At first sight, it seems strange that there exists such a solution 
because the fermion determinant in the partition function \eqref{Zalphabyeigenvalue} looks 
\begin{align} 
\prod_{i,j=1}^N\left(e^{i\alpha}-1+\lambda_i+\lambda_j\right)
=\prod_{i=1}^N\left(e^{i\alpha}-1+2\lambda_i\right) \,
\prod_{i>j}\left(e^{i\alpha}-1+\lambda_i+\lambda_j\right)^2,
\label{fermiondet}
\end{align} 
the first factor of which makes $\lambda_i$ apart from the origin in the $\alpha\rightarrow 0$ limit. 
However, as we will see below, the symmetric one-cut solution exists owing to the large-$N$ limit 
and we will confirm its validity by checking finiteness of the free energy for our solution.  

In order to solve \eqref{DWSPE} for $\rho(x)$ with a symmetric support $[-c,c]$, 
let us consider a complex function 
\begin{align}
G(z)\equiv \int_{-c}^cdy\frac{\rho(y)}{z-y},  
\end{align}
and further define as in \cite{Eynard:1995nv}
\begin{align}
G_-(z)\equiv\frac12(G(z)-G(-z)),
\end{align}
then $G_-(z)$ has following properties: 
\begin{enumerate}
\item $G_-(z)$ is odd, analytic in $z\in {\bf C}$ except the cut $[-c,c]$.
\item $G_-(x)\in {\bf R}$ for $x\in {\bf R}$ and $x\notin [-c,c]$.
\item $G_-(z)\rightarrow \frac{1}{z} +\cO(\frac{1}{z^3})$ as $z\rightarrow\infty$.
\item $G_-(x\pm i0)=\frac{1}{2}(x^2-\mu^2)x\mp i\pi\rho(x)$ for $x\in[-c,c]$. 
\end{enumerate}
They lead us to deduce 
\begin{align}
G_-(z)=\frac12(z^2-\mu^2)z-\frac12\left(z^2-\mu^2+\frac{c^2}{2}\right)\sqrt{z^2-c^2}
\end{align}
with
\begin{align}
c^2(3c^2-4\mu^2)=16,
\end{align}
from which we find that 
\begin{align}
\rho_0(x)=\frac{1}{2\pi}\left(x^2-\mu^2+\frac{c^2}{2}\right)\sqrt{c^2-x^2},~~~x\in[-c,c].
\label{symmonecut}
\end{align}
$\rho_0(x)\geq 0$ tells us that this solution is valid for $\mu^2\leq 2$, 
which is indeed the complement of the region of $\mu^2$ 
where both the two-cut solution and the asymmetric one-cut solution obtained in \cite{Kuroki:2009yg} exist. 
Given $\rho_0(x)$, it is straightforward to calculate the free energy $\eqref{F0}$ as 
\begin{align}
F_0(x)=\frac13x^2-\frac{1}{216}x^4-\frac{1}{216}(x^3+30x)\sqrt{x^2+12}-\log(x+\sqrt{x^2+12})+\log 6,
\end{align}
where $x=\mu^2$. In contrast to this, it is observed in \cite{Kuroki:2009yg} that 
for $\mu^2>2$ the free energy calculated from the asymmetric one-cut solution 
or the two-cut solution is independent of $\mu^2$ and vanishes, reflecting 
the restoration of the SUSY. It is easy to see that $F_0(x)>0$ for $x<2$, and 
the expectation value of $\frac{1}{N}\tr B$ is computed to be 
\be
\vev{\frac{1}{N} \tr B}=-i\left[\frac{c^4}{16}(c^2-x)-x\right],
\ee
which is nonzero for $x<2$. 
These are strong evidence suggesting the spontaneous SUSY breaking. 
Also, the $x$-derivatives of the free energy,  
\be
\lim_{x\to 2-0}F_0(x)=\lim_{x\to 2-0}\frac{dF_0(x)}{dx}=\lim_{x\to 2-0}\frac{d^2F_0(x)}{dx^2}=0, 
\qquad \lim_{x\to 2-0}\frac{d^3F_0(x)}{dx^3}=-\frac{1}{2},
\ee 
show that the transition between the SUSY phase ($x\geq 2$) and the SUSY broken phase ($x<2$) is of 
the third order. 

As commented in \eqref{fermiondet}, if we take a look at the ${\cal O}(1/N)$ contribution 
of $F$ given in (\ref{F_1}), 
naively it seems strange that we have a nonzero saddle point eigenvalue density 
around the origin, since $xW^{\prime\prime}(x)=2x^2\sim 0$ there and the integrand of (\ref{F_1}) diverges. 
Furthermore, it is also curious that we have eigenvalues in general distributed 
around zeros of $W'(x)$, because the integrand again diverges there 
which would mean that the partition function vanishes. 
However notice that, even if the integrand looks divergent, 
it is just logarithmic and its integral itself is finite due to a contribution from the measure. 
Because $\rho_0(x)$ is finite, the relevant integral over the vicinity of the singularities 
for the real part~\footnote{The imaginary part is 
irrelevant in the analysis, because it just 
contributes to the overall sign of the partition function as 
${\rm sgn}\left(\prod_{i=1}^N\left(W''(x_i)W'(x_i)x_i\right)\right)$ 
in evaluating the partition function at a single large-$N$ solution.}    
 of $F_1$ is at most 
$
\int_0^\varepsilon dx\, \log x
$ 
which clearly converges at the origin. 
(The logarithmic singularity is integrable.) 
Therefore, ${\cal O}(1/N)$ part of $F$ does not diverge owing to the large-$N$ limit. 

As an example, in the double-well case $W'(\phi)=\phi^2-\mu^2$ with $\mu^2>2$, 
let us consider the two solutions obtained in \cite{Kuroki:2009yg}. The asymmetric one-cut solution is given by  
\begin{align}
\rho_0(x)=\frac{x}{\pi}\sqrt{(x^2-a^2)(b^2-x^2)},~~~a^2=-2+\mu^2,~~~b^2=2+\mu^2,
\end{align}
and the two-cut solution with the filling fraction ($\nu_+, \nu_-$) is (\ref{two-cut_sol}). 
Evaluating (\ref{F_1}) at each of the solutions, we find the same result of the real part of $F_1$ for both solutions : 
\begin{align}
{\rm Re}\,F_1= -\log\left(\mu^2+\sqrt{\mu^4-4}\right) +1-\frac14\mu^2\left(\mu^2-\sqrt{\mu^4-4}\right). 
\end{align}
It is finite as understood from the above reasoning, and interestingly it is not dependent on $\nu_\pm$. 
The finiteness supports the validity of our large-$N$ solutions. 

Finally, we make a comment on one of interesting aspects of our result that 
the existence of a leading nontrivial contribution at ${\cal O}\left((e^{i\alpha}-1)^N\right)$ 
given in \eqref{Zalphabyrho} 
suggests SUSY breaking for finite $N$, 
but that in the double-well case (\ref{DW}) with $\mu^2> 2$ 
it leads to the supersymmetric solutions to \eqref{SPE} 
in the large-$N$ limit with vanishing free energy, as discussed below (\ref{alpha=0}).

\section{Summary and discussion}
\label{sec:summary}
\setcounter{equation}{0}
In this paper, firstly we discussed localization in discretized SUSY quantum mechanics 
without the external field $\alpha$ 
by changing integration variables. It makes it clear that the path integral is localized 
at the auxiliary field $B=0$, which in turn implies the standard localization 
at the critical points of superpotential.  
Furthermore, it was investigated in detail how $\Xi_0(B)$, 
the integrand of the partition function with respect to $B$, behaves as $B\sim 0$, and 
clarified whether the change of variables is applicable or not. 
Similar arguments were presented also for $\alpha\neq 0$ case. 
This gives a different approach to localization from a deformation by $Q$-exact terms. 
It is worth pointing out that the change of variables can be applied even to systems 
where SUSY is (explicitly) broken, while a deformation by $Q$-exact terms cannot be straightforwardly. 
Thus, the former is useful to investigate localization in systems where the external field explicitly 
breaking SUSY is turned on. We also stress that we provided a firm formulation 
of change of variables for localization (without issues mentioned in footnote \ref{footnote}) 
in the path integral which is useful in discussion of spontaneous 
SUSY breaking. As emphasized in the introduction, such a formulation is indispensable 
because nonperturbative formulations of string theory in terms of matrix models 
are defined by the path integral.   

Secondly, we explained localization in SUSY matrix models without the external field. 
The formula of the partition function was obtained, which is given by the $N$-th power of the localization 
formula in the $N=1$ case ($N$ is the rank of matrix variables). 
It can be regarded as a matrix-model generalization of the ordinary localization formula.     
In terms of eigenvalues, localization attracts them to the critical points of superpotential, 
while the square of the Vandermonde determinant originating from the measure factor 
gives repulsive force among them. 
Thus, the dynamics of the eigenvalues is governed by balance of the attractive force 
from the localization and the repulsive force from the Vandermonde determinant. 
It is a new feature specific to SUSY matrix models, not seen 
in the discretized SUSY quantum mechanics. 
For a general superpotential which has $K$ critical points, contribution to the partition function 
from $\nu_IN$ eigenvalues fluctuating around the $I$-th critical point ($I=1, \cdots, K$), denoted by 
$Z_{(\nu_1, \cdots,\nu_K)}$, was shown to be equal to the products of the partition functions of the 
Gaussian SUSY matrix models $Z_{G,\nu_1}\cdots Z_{G, \nu_K}$. 
Here, $Z_{G, \nu_I}$ is the partition function of the Gaussian SUSY matrix model 
with $\nu_IN$ the rank of matrix variables, which describes Gaussian fluctuations 
around the $I$-th critical point.  
In the double-well case, it leads to the claim 
of the previous paper~\cite{Kuroki:2009yg}. 
It is interesting to investigate whether such a factorization occurs also 
for various expectation values.      

Thirdly, as mentioned in the above, the argument of the change of variables 
leading to localization can be applied to $\alpha\neq 0$ case. 
Then, we found that $\alpha$-dependent terms in the action explicitly break SUSY 
and makes localization incomplete. 
Instead of it, the Nicolai mapping, which is also applicable to the $\alpha\neq 0$ case, 
is more convenient for actual calculation in SUSY matrix models. 
In the case that the supersymmetric partition function (the partition function with $\alpha=0$) vanishes,    
we obtained an exact result of a leading nontrivial contribution to the partition function 
with $\alpha\neq 0$ in the expansion of $(e^{i\alpha}-1)$ for finite $N$. 
It will play a crucial role to compute various correlators when SUSY is spontaneously broken. 
Large-$N$ solutions for the double-well case $W'(\phi)=\phi^2-\mu^2$ were derived, 
and it was found that there is a phase transition between the SUSY phase 
corresponding to $\mu^2>2$ and the SUSY broken phase to $\mu^2<2$. 
It was shown to be of the third order.  

For future directions, 
this kind of argument can be expected to be useful to investigate localization in various lattice models 
for supersymmetric field theories which realize some SUSYs on the lattice~\footnote{
For examples of such lattice formulations, 
see \cite{Sakai:1983dg,Kaplan_review,Kaplan:2002wv,Catterall:2003wd,Catterall:2004np,SYM_lat,Endres:2006ic,SQCD_lat,Ishii:2008ib,Hanada:2010kt,D'Adda:2010pg}.}.   

Also, it will be interesting to investigate localization in models constructed in ref.~\cite{Kuroki:2007iy}, which couple a supersymmetric quantum field theory to a certain large-$N$ matrix model 
and cause spontaneous SUSY breaking at large $N$.   

Finally, we hope that similar analysis for super Yang-Mills matrix 
models~\cite{Banks:1996vh,Ishibashi:1996xs,Dijkgraaf:1997vv}, which have been proposed as nonperturbative 
definitions of superstring/M theories,  
will shed light on new aspects of spontaneous SUSY breaking in superstring/M theories.  
To carry out the analysis, the method of the Gaussian expansion or improved perturbation theory would be 
useful~\cite{Oda:2000im,Nishimura:2001sx,Nishimura:2002va,Aoyama:2006yx,Aoyama:2010ry}.

\section*{Acknowledgements}
We would like to thank High Energy Accelerator Research Organization (KEK) and 
Yukawa Institute for Theoretical Physics at Kyoto University. Discussions during the KEK theory workshop (March, 2010) and 
the YITP workshop YITP-W-10-02 on ``Development of Quantum Field Theory and String Theory'' (July, 2010) were 
useful to complete this work.   
F.~S. is grateful to Tetsuo~Horigane, Daisuke~Kadoh, Hiroki~Kawai, Yoshio~Kikukawa and Hiroshi~Suzuki 
for enjoyable discussions on supersymmetric theories. 
Also, F.~S. appreciates the Aspen Center for Physics where a part of this work was done during his visit (May and June, 2010). 
The work of T.~K. is supported in part by Rikkyo University Special Fund for Research, 
and the work of F.~S. is supported in part by a Grant-in-Aid for Scientific Research (C), 21540290.

\appendix
\section{Localization in discretized SUSY quantum mechanics with general $T$}
\label{app:loc_generalT}
In this appendix, we generalize the argument for $T=1$ 
presented in sections~\ref{subsec:Loc_in_QM} and \ref{subsec:Loc_in_QM2} to $T\ge 2$ case. 

\subsection{Localization for general $T$} 
The action (\ref{dQM_S2}) with the periodic boundary condition ($\alpha=0$), denoted by $S_0$, 
is invariant under the $Q$-SUSY: 
\be
Q\phi(t)=\psi(t), \quad Q\psi(t)=0, \quad Q\bar{\psi}(t)=-iB(t), \quad QB(t)=0.
\ee
We consider the field redefinition 
\be
\phi(t)=\tilde{\phi}(t) + \bar\epsilon\psi(t), \qquad \bar\psi(t)=\tilde{\bar{\psi}}(t) -i\bar\epsilon B(t), 
\label{fieldredefT}
\ee
where $\tilde{\bar{\psi}}(t)$ is chosen to be orthogonal to $-i\bar\epsilon B(t)$ as  
\be
\sum_{t=1}^TB(t) \tilde{\bar{\psi}}(t)=0.
\label{ortho_tilde_bar_psi_T}
\ee
Namely, the degrees of freedom of $\bar{\psi}(t)$ are carried by both 
$\bar{\epsilon}$ and $\tilde{\bar{\psi}}(t)$. 
Differently from the $T=1$ case, $\bar\epsilon$ cannot parametrize 
the whole functional space of $\bar\psi(t)$, and represents merely a single degree of freedom 
of $\bar\psi(t)$ parallel to $B(t)$.

Similarly to the $T=1$ case, since (\ref{fieldredefT}) is recast as a shift by the $Q$ transformation:  
\be
\phi(t) = \tilde{\phi}(t)+\bar\epsilon Q\tilde{\phi}(t), \qquad 
\bar\psi(t) = \tilde{\bar{\psi}}(t)+\bar\epsilon Q\tilde{\bar{\psi}}(t),
\ee 
the action $S_0$ is shown to be independent of $\bar\epsilon$: 
\bea
S_0(B,\phi,\psi,\bar{\psi}) & = & S_0(B+\bar{\epsilon}QB,\,\tilde{\phi} + \bar\epsilon Q\tilde{\phi}, \,
\psi + \bar\epsilon Q\psi,\,\tilde{\bar\psi}+\bar\epsilon Q\tilde{\bar\psi}) \nn \\
 & = & S_0(B, \tilde{\phi}, \psi,\tilde{\bar\psi}) \nn \\
 & = & \sum_{t=1}^T\left[\frac12 B(t)^2 +iB(t)\left\{\tilde{\phi}(t+1)-\tilde{\phi}(t)
+W'(\tilde{\phi}(t))\right\} \right. \nn \\
& & \hspace{7mm} 
\left. \frac{}{} +\tilde{\bar\psi}(t)\left\{ \psi(t+1)-\psi(t)+W''(\tilde{\phi}(t))\psi(t)\right\}
\right],
\label{S0_fieldredefT}
\eea
from the $Q$-SUSY invariance of $S_0$.   

When we write the partition function as 
\bea
Z_0 & = & \int \prod_{t=1}^T dB(t)\, \Xi_0(B), \\
\Xi_0(B) & \equiv & \left(\frac{-1}{2\pi}\right)^T 
\int \prod_{t=1}^T\left(d\phi(t)\,d\psi(t)\,d\bar\psi(t)\right) \,
e^{-S_0} \nn \\
 & = & (-1)^{T(T-1)/2}\int \prod_{t=1}^T d\psi(t) \left(\frac{-1}{2\pi}\right)^T 
 \int \prod_{t=1}^T\left(d\phi(t)\,d\bar\psi(t)\right) \,e^{-S_0},  
\eea
we can regard $B(t)$ and $\psi(t)$ as external fields in the integral 
$\int \left(\prod_{t=1}^Td\phi(t)\,d\bar\psi(t)\right) \,e^{-S_0}$. 
This point of view makes easier to derive 
the Jacobian of the path-integral measure associated with (\ref{fieldredefT}). 
For $-i\bar\epsilon B(t)$ in (\ref{fieldredefT}), we decompose it as 
\be
-i\bar\epsilon B(t) = -i \cN_B\bar\epsilon \times \frac{1}{\cN_B} B(t),
\ee
with $\cN_B\equiv ||B||\equiv \sqrt{\sum_{t=1}^TB(t)^2}$. Since the second factor $\frac{1}{\cN_B} B(t)$ can be 
regarded as ``a normalized wave function'', i.e. a unit vector in the functional space of $B(t)$, 
the remaining $-i \cN_B\bar\epsilon$ is identified with 
an integration variable. Thus, the measure of $\bar\epsilon$ associated with \eqref{fieldredefT} 
is given by $d(-i\cN_B\bar{\epsilon})= \frac{i}{\cN_B}d\bar\epsilon$. 
Also for $\tilde{\bar\psi}(t)$, expressing the constraint (\ref{ortho_tilde_bar_psi_T}) as a delta-function, 
the measure is explicitly defined by~\footnote{
The sign factor $(-1)^{T-1}$ can be determined so that the RHS becomes $\prod_{t=2}^T d\tilde{\bar\psi}(t)$ 
when $B(t)$ vanishes except $B(1)$.}   
\bea
\left(d\tilde{\bar\psi}\right) &\equiv &(-1)^{T-1}\Bigl(\prod_{t=1}^T d\tilde{\bar\psi}(t) \Bigr)\,
\delta\left(\frac{1}{\cN_B}\sum_{t=1}^TB(t)\tilde{\bar\psi}(t)\right) \nn \\
& = & (-1)^{T-1}\Bigl(\prod_{t=1}^T d\tilde{\bar\psi}(t)\Bigr) \,
\frac{1}{\cN_B}\sum_{t=1}^TB(t)\tilde{\bar\psi}(t). 
\label{measure_psibarT}
\eea
Hence, we obtain the measure for $\bar\psi$ as~\footnote{
For $T=1$, $\left(d\tilde{\bar\psi}\right)$ is reduced to $\frac{B(1)}{\cN_B}$, so we have 
$
d\bar\psi(1)=\frac{i}{\cN_B^2}\,B(1) d\bar\epsilon = \frac{i}{B(1)}\,d\bar\epsilon
$
which reproduces the Jacobian in (\ref{QMJacobian}).
} 
\be
\prod_{t=1}^T d\bar\psi(t) = \frac{i}{\cN_B}d\bar\epsilon\,\left(d\tilde{\bar\psi}\right).
\label{d_psi_barT}
\ee

After the change of variables (\ref{fieldredefT}), $\Xi_0(B)$ becomes 
\bea
\Xi_0(B) & = &\frac{(-1)^{T(T+1)/2}}{(2\pi)^T}\frac{i}{\cN_B}\,e^{-\sum_{t=1}^T\frac12B(t)^2}
 \left(\int \prod_{t=1}^T d\tilde\phi(t)\,e^{-i\sum_{t=1}^TB(t)\left\{\tilde{\phi}(t+1)-\tilde{\phi}(t) 
+W'(\tilde{\phi}(t))\right\}}\right) \nn \\
 & & \times \int \prod_{t=1}^T d\psi(t)\int d\bar\epsilon\,\left(d\tilde{\bar\psi}\right) \,
e^{-\sum_{t=1}^T\tilde{\bar\psi}(t)\left\{\psi(t+1)-\psi(t)+W''(\phi(t))\psi(t)\right\}}.
\label{Xi0_fieldredefT}
\eea
It is clear that $\Xi_0(B)$ vanishes due to the trivial $\bar\epsilon$-integral as long as $\cN_B\neq 0$. 
Thus, the path integral of the partition function is localized to $\cN_B=0$, 
i.e. $B(1)=\cdots =B(T)=0$.    

On the other hand, as in \eqref{Z_0^0_fieldredef} of the $T=1$ case, 
the change of variables \eqref{fieldredefT} 
does not work when $\cN_B\sim 0$. In order to see how $\Xi_0(B)$ becomes singular 
as $\cN_B \to 0$, let us consider (\ref{Xi0_fieldredefT}) in the 
region $\cN_B<\varepsilon$. 
We express $(B(1),\cdots, B(T))\in {\bf R}^T$ by polar coordinates with the radial direction 
$\cN_B$ and the angular directions specified by the unit vector  
\be
\Omega_B(t)\equiv \frac{1}{\cN_B}\,B(t). 
\ee 
For $W'(\phi)$ given by (\ref{W'}) with $p\geq 2$, we rescale $\tilde{\phi}(t)$ and $\tilde{\bar\psi}(t)$ as 
\be
\tilde{\phi}(t) = \left(\frac{1}{\cN_B}\right)^{\frac{1}{p}}\,\phi'(t), \qquad 
\tilde{\bar\psi}(t)=\cN_B^{\frac{p-1}{p}}\,\bar{\psi}'(t), 
\ee
which correspondingly changes the measure as  
\be
\prod_{t=1}^T d\tilde{\phi}(t) = \left(\frac{1}{\cN_B}\right)^{\frac{T}{p}}\, \prod_{t=1}^T d\phi'(t) , \qquad 
\left(d\tilde{\bar\psi}\right) = \left(\frac{1}{\cN_B}\right)^{\frac{p-1}{p}(T-1)}\,\left(d\bar\psi'\right). 
\ee
The rescaling is convenient to see the $\cN_B$-dependence of $\Xi_0(B)$. 
Then, the integrands of the $\tilde{\phi}$-integral and the Grassmann integral become   
\bea
& & e^{-i\sum_{t=1}^TB(t)\left\{\tilde{\phi}(t+1)-\tilde{\phi}(t) +W'(\tilde{\phi}(t))\right\}}
= e^{-i\sum_{t=1}^T\Omega_B(t)g_p\phi'(t)^p}\,\left[1+\cO(\varepsilon^{1/p})\right], \nn \\
& & e^{-\sum_{t=1}^T\tilde{\bar\psi}(t)\left\{\psi(t+1)-\psi(t)+W''(\tilde{\phi}(t))\psi(t)\right\}}
= e^{-\sum_{t=1}^T\bar\psi'(t)pg_p\phi'(t)^{p-1}\psi(t)}\,\left[1+\cO(\varepsilon^{1/p})\right]
\nn \\ 
\eea
for $\cN_B < \varepsilon$, respectively. Plugging the above results, $\Xi_0(B)$ can be expressed as 
\bea
\Xi_0(B) & = & i\frac{(-1)^{T(T+1)/2}}{(2\pi)^T}\,\left(\frac{1}{\cN_B}\right)^{T+\frac{1}{p}} e^{-\frac12\cN_B^2}
\int\prod_{t=1}^T d\phi'(t)\,e^{-i\sum_{t=1}^T\Omega_B(t) g_p\phi'(t)^p} \nn \\
& & \times \int \prod_{t=1}^T d\psi(t) \int d\bar\epsilon \left(d\bar\psi'\right) \,
e^{-\sum_{t=1}^T\bar\psi'(t)pg_p\phi'(t)^{p-1}\psi(t)}\,\left[1+\cO(\varepsilon^{1/p})\right]. 
\label{Xi0_fieldredefT2}
\eea
We thus find that $\Xi_0(B)$ becomes singular as $\cN_B^{-T-\frac{1}{p}}$ for $\cN_B\sim 0$. 

More precisely, if we define $Z_0^{(0)}=\int_{\cN_B<\varepsilon} \prod_{t=1}^T dB(t)\, \Xi_0(B)$ 
as in \eqref{decomposition}, the factor $\cN_B^{\,T-1}$ coming from the measure 
$\prod_{t=1}^T dB(t)=\cN_B^{\,T-1} d\cN_B d\Omega_B$ expressed in the polar coordinates makes somewhat milder 
the singularity at $\cN_B=0$ in $\Xi_0(B)$. However, it is not sufficient to achieve convergence because   
\be
\int^\varepsilon_0d\cN_B \, \cN_B^{\,T-1}\,\left(\frac{1}{\cN_B}\right)^{T+\frac{1}{p}} e^{-\frac12\cN_B^2}
= \int^\varepsilon_0d\cN_B \,\left(\frac{1}{\cN_B}\right)^{1+\frac{1}{p}}
\times \left[1+\cO(\varepsilon^2)\right]=\infty.
\label{Bintnearorigin}
\ee   
Hence, in $Z_0^{(0)}$, the trivial $\bar\epsilon$-integral vanishes while the $\cN_B$-integral diverges. 
We see that the change of variables (\ref{fieldredefT}) is not appropriate for $Z_0^{(0)}$, 
because it leads to an indefinite expression: 
$Z_0^{(0)}=\infty \times 0$. 
(In $p=1$ case, we can consider $W'(\phi)=g_1\phi$, 
because the constant term $g_0$ can be absorbed by a shift of $\phi$. 
The $\tilde{\phi}$-integrals in (\ref{Xi0_fieldredefT}) yield 
\bea
& & \frac{1}{(2\pi)^T}\int \prod_{t=1}^T d\tilde\phi(t)\,
e^{-i\sum_{t=1}^TB(t)\left\{\tilde{\phi}(t+1)-\tilde{\phi}(t) +W'(\tilde{\phi}(t))\right\}} \nn \\
& & = 
\frac{1}{(2\pi)^T}\int \prod_{t=1}^T d\tilde\phi(t)\,
e^{i\sum_{t=1}^T\left\{(1-g_1)B(t)-B(t-1)\right\}\tilde{\phi}(t)} \nn \\
& & = \prod_{t=1}^T\delta \left((1-g_1)B(t)-B(t-1)\right) = \frac{1}{|1-(1-g_1)^T|}\,\prod_{t=1}^T 
\delta(B(t))
\eea
with $B(0)\equiv B(T)$. Due to the delta-functions $\prod_{t=1}^T \delta(B(t))$ and 
$1/\cN_B$ in \eqref{Xi0_fieldredefT}, the $B$-integrals 
in $Z_0^{(0)}$ become singular leading to an indefinite form of $Z_0^{(0)}$: $\infty \times 0$. )  
As in the $T=1$ case, the indefinite form of $Z^{(0)}_0$ obtained 
after the change of variables (\ref{fieldredefT}) implies that $Z^{(0)}_0$ possibly takes a nontrivial value.

\paragraph{Unnormalized expectation values} 
For the unnormalized expectation values of $B(t)^n$ ($n\geq 1$): 
\be
\vev{B(t)^n}'\equiv \int \Bigl(\prod_{t=1}^T dB(t)\Bigr)\, B(t)^n\,\Xi_0(B), 
\ee
by the same change of variables, the contribution from the region $\cN_B=||B||> \varepsilon$ clearly 
vanish from $\int d\bar\epsilon =0$. Hence,  
\bea
\vev{B(t)^n}' & = & \frac{(-1)^{T(T+1)/2}}{(2\pi)^T} \,i\int_{||B||<\varepsilon} \Bigl(\prod_{t=1}^TdB(t)\Bigr)\,\frac{B(t)^n}{\cN_B} 
\int \prod_{t=1}^T\left(d\tilde\phi(t)\,d\psi(t)\right) \nn \\
 & & \times \int d\bar\epsilon \, \left(d\tilde{\bar\psi}\right) \,
e^{-S_0(B, \tilde{\phi}, \psi,\tilde{\bar\psi})}. 
\label{Bn_unVEV_fieldredefT}
\eea
If $W'(\phi)$ is given by (\ref{W'}) with $p\geq 2$, using (\ref{Xi0_fieldredefT2}) 
and the polar coordinates for $(B(1), \cdots, B(T))$, we have
\bea
\vev{B(t)^n}' & = & i
\left(\int^\varepsilon_0 d\cN_B \,\cN_B^{\,n-1-\frac{1}{p}}e^{-\frac12\cN_B^2}\right) \,
Y_n(t)\left[1+\cO(\varepsilon^{1/p})\right], 
\label{Bn_unVEVT_int}\\
Y_n(t)& \equiv &  \frac{(-1)^{T(T+1)/2}}{(2\pi)^T}\,
\int d\Omega_B\,\Omega_B(t)^n 
\int\prod_{t=1}^T d\phi'(t)\,e^{-i\sum_{t=1}^T\Omega_B(t) g_p\phi'(t)^p} \nn \\
& & \times \int \prod_{t=1}^T d\psi(t) \int d\bar\epsilon \left(d\bar\psi'\right) \,
e^{-\sum_{t=1}^T\bar\psi'(t)pg_p\phi'(t)^{p-1}\psi(t)}. 
\label{Yn}
\eea
Note that, since the $\cN_B$-integral is not singular at the origin for $n\geq 1$, 
the field redefinition (\ref{fieldredefT}) is always possible, 
differently from the case of the partition function. 
(For completeness, we show that $Y_n(t)$ definitely vanishes, 
i.e. factors in front of trivial Grassmann integrals are finite (in fact, they vanish)   
in appendix~\ref{app:Yn}.) 
Thus, we can show 
\be
\vev{B(t)^n}'=0 \qquad (n\geq 1),
\label{Bn_unVEVT0}
\ee 
due to the trivial $\bar\epsilon$-integral. 
In $p=1$ case, the delta-functions $\prod_{t=1}^T\delta (B(t))$ arise after the $\tilde{\phi}$ integration 
in (\ref{Bn_unVEV_fieldredefT}), which makes the $B$-integrals finite for $n\geq 1$ as~\footnote{
The result (\ref{Bn_unVEVT_p=1}) is obtained, if we integrate $B(t')$ (${}^\forall t'\neq t$) 
before the $B(t)$-integral. Otherwise, we would have the vanishing result even for $n=1$. 
We choose the order so that the result is reduced to (\ref{Bn_unVEV_p=1}) when $T=1$.
}   
\be
\int_{||B||<\varepsilon}\Bigl(\prod_{t=1}^T dB(t)\Bigr) \, \frac{B(t)^n}{\cN_B}\, e^{-\frac12\cN_B^2}
\,\prod_{t=1}^T\delta (B(t))=\delta_{n, 1}. 
\label{Bn_unVEVT_p=1}
\ee
It leads to (\ref{Bn_unVEVT0}) from the trivial $\bar\epsilon$-integral. 
   
In general, we find that the unnormalized expectation values 
of $B(1)^{n_1}\cdots B(T)^{n_T}$ with $n_1, \cdots, n_T=0,1,2,\cdots$ and $\sum_{t=1}^T n_t\geq 1$ 
vanish: 
\be
\vev{B(1)^{n_1}\cdots B(T)^{n_T}}'=0. 
\label{Bn_unVEV_T}
\ee
 
\paragraph{Localization to $\phi(t+1)-\phi(t)+W'(\phi(t))=0$}
Because 
\be
\vev{e^{-\frac{u-1}{2}\sum_{t=1}^TB(t)^2}}'
=\sum_{n=0}^\infty \frac{1}{n!}\left(-\frac{u-1}{2}\right)^n\vev{\left(\sum_{t=1}^T B(t)^2\right)^n}'
=\vev{1}'=Z_0 
\ee
holds for an arbitrary parameter $u$ from (\ref{Bn_unVEV_T}), 
the partition function can be computed similarly to the $T=1$ case. 
Taking $u>0$ and integrating with respect to $B$, we have~\footnote{
An explicit computation of $Z_0$ is given in appendix A in~\cite{Giedt:2004qs}, where deformation invariance 
by $Q$-exact terms is used to obtain 
\be
Z_0 = \left(\sum_{\phi:\,W'(\phi)=0} \frac{W''(\phi)}{|W''(\phi)|}\right)^T = \sharp^T.
\label{localization_T}
\ee
It is the $T$-th power of the result of the $T=1$ case.
}   
\bea
Z_0 & = & \int \prod_{t=1}^Td\phi(t)\,\frac{1}{(2\pi u)^{T/2}}\,
e^{-\frac{1}{2u}\sum_{t=1}^T\left(\phi(t+1)-\phi(t)+W'(\phi(t))\right)^2} \nn \\
  & & \times (-1)^T\int \Bigl(\prod_{t=1}^Td\psi(t)\,d\bar\psi(t)\Bigr)\,
e^{-\sum_{t=1}^T\bar\psi(t)\left\{\psi(t+1)-\psi(t)+W''(\phi(t))\psi(t)\right\}}.
\eea
In the limit $u\to 0$, the integration with respect to $\phi(t)$ is manifestly localized 
to configurations satisfying $\phi(t+1)-\phi(t)+W'(\phi(t))=0$.

\subsection{Localization in the presence of external field}
The action (\ref{dQM_S2}) with the twisted boundary condition (\ref{TBC_alpha}), 
denoted by $S_\alpha$, can be written as 
\be
S_\alpha = S_{0} + (e^{i\alpha}-1)\bar{\psi}(T)\psi(1), 
\ee
where $S_0$ is the action with the periodic boundary condition. 
As mentioned below (\ref{TBC_alpha}), $\phi(T+1)$ and $\psi(T+1)$ were replaced with 
$\phi(1)$ and $e^{i\alpha}\psi(1)$, respectively. 
Under the change of variables (\ref{fieldredefT}) which is defined for variables at $t=1, \cdots, T$, 
this becomes 
\be
S_\alpha (B, \phi, \psi, \bar\psi) = S_0(B, \tilde{\phi}, \psi, \tilde{\bar\psi}) + 
(e^{i\alpha}-1)\tilde{\bar\psi}(T)\psi(1)-i(e^{i\alpha}-1)\bar\epsilon B(T)\psi(1).
\ee 
The first term is given by (\ref{S0_fieldredefT}), independent of $\bar\epsilon$. 
Due to the last term $-i(e^{i\alpha}-1)\bar\epsilon B(T)\psi(1)$, 
the $\bar\epsilon$-integral in the partition function does not vanish.

Similarly to the $\alpha=0$ case, let us write the partition function as 
\bea
Z_\alpha & = & \int \prod_{t=1}^T dB(t)\, \Xi_\alpha(B), \\
\Xi_\alpha(B) & \equiv & \left(\frac{-1}{2\pi}\right)^T 
\int \prod_{t=1}^T\left(d\phi(t)\,d\psi(t)\,d\bar\psi(t)\right) \,
e^{-S_\alpha} \nn \\
 & = & (-1)^{T(T-1)/2}\int \prod_{t=1}^T d\psi(t) \left(\frac{-1}{2\pi}\right)^T 
 \int \prod_{t=1}^T\left(d\phi(t)\,d\bar\psi(t)\right) \,e^{-S_\alpha}, 
\eea
also 
\bea
& & Z_\alpha =  Z_\alpha^{(0)} + \tilde{Z}_\alpha, \nn \\
& & Z^{(0)}_\alpha = \int_{||B||<\varepsilon} \prod_{t=1}^T dB(t) \, \Xi_\alpha(B), \qquad
\tilde{Z}_\alpha = \int_{||B||\geq\varepsilon} \prod_{t=1}^T dB(t) \, \Xi_\alpha(B).
\eea
The field redefinition (\ref{fieldredefT}) recasts $\Xi_\alpha(B)$ to 
\bea
\Xi_\alpha(B) &  = &(e^{i\alpha}-1)\,\frac{B(T)}{\cN_B}\,\frac{(-1)^{T(T-1)/2}}{(2\pi)^T} 
\int \Bigl(\prod_{t=1}^T d\tilde{\phi}(t)\Bigr) \nn \\
& & \hspace{21mm}\times
\int \Bigl(\prod_{t=2}^T d\psi(t)\Bigr)\,\left(d\tilde{\bar\psi}\right)
\,e^{-\left.S_0(B, \tilde{\phi}, \psi, \tilde{\bar\psi})\right|_{\psi(1)=0}}, 
\eea 
after integrating over $\bar\epsilon$ and $\psi(1)$. 
In the process, we used 
\bea
\int d\psi(1)\, d\bar\epsilon\, e^{i(e^{i\alpha}-1)\bar{\epsilon} B(T)\psi(1)} 
& = & \int d\psi(1)\, d\bar\epsilon\,\left[1+i(e^{i\alpha}-1)\bar{\epsilon} B(T)\psi(1)\right] \nn \\
& = & i(e^{i\alpha}-1)B(T). 
\label{psi(1)_barepsilon_int}
\eea
Note that it is valid for $\cN_B=||B|| \neq 0$. 
In the case $\cN_B\sim 0$, we should keep the first term ``1" in the expansion of 
$e^{i(e^{i\alpha}-1)\bar{\epsilon} B(T)\psi(1)}$ in the RHS of the first equality 
in (\ref{psi(1)_barepsilon_int}), 
although it gives the vanishing $\psi(1)$- and $\bar\epsilon$-integrals. 
Integrating it over $||B||<\varepsilon$ yields a singularity at the origin, 
so we have an indefinite form ($\infty \times 0$) 
which cannot be discarded safely.  It is parallel to the situation of the $T=1$ case discussed in 
footnote~\ref{fn:T=1}.  

The contribution to the partition function from the integration region $||B||\geq \varepsilon$: 
$\tilde{Z}_\alpha$ is in general nonvanishing. 
However, when $W'(\phi)$ is linear, the $\tilde{\phi}$-integrals yield $\prod_{t=1}^T\delta(B(t))$ leading to 
$\tilde{Z}_\alpha=0$.  

On the other hand, the partition function can be computed directly from (\ref{Z_alpha_T}) 
without using (\ref{fieldredefT}). Since the fermion determinant can be written as the sum of 
the determinant under the periodic boundary condition ($\prod_{t=1}^T(-1+W''(\phi(t))) -(-1)^T$) and 
the effect of the twist ($-(-1)^T(e^{i\alpha}-1)$), we have 
\be
Z_\alpha  =  Z_0 -(e^{i\alpha}-1)\left(\frac{-1}{\sqrt{2\pi}}\right)^T 
\int \Bigl(\prod_{t=1}^Td\phi(t)\Bigr)\,e^{-\frac12 \sum_{t=1}^T\left\{\phi(t+1)-\phi(t)+W'(\phi(t))\right\}^2}.
\label{Z_alpha_Z_0_T}
\ee
Since the second term is the net effect of the twist, it should be equal to 
the sum of $Z_\alpha^{(0)}-Z_0^{(0)}$ and $\tilde{Z}_\alpha$. 
Note again that although it vanishes in the $\alpha\rightarrow 0$ limit, 
it becomes important when the SUSY is spontaneously broken, i.e. $Z_0=0$. 
Let us elaborate on the former as in the $T=1$ case. 
When $W'(\phi)$ is linear ($W'(\phi)=g_1\phi+g_0$), 
we explicitly obtain 
\be
Z_\alpha^{(0)}-Z_0^{(0)} = -(e^{i\alpha}-1)\,\frac{(-1)^T}{|1-(1-g_1)^T|}, \qquad 
\tilde{Z}_\alpha = 0, 
\ee
which again means that the localization takes place even in the presence of the external field, 
and that the effect of the twist on $Z_\alpha^{(0)}$ remains 
even in the $\varepsilon\rightarrow 0$ limit. 
It can be understood from $\Xi_\alpha(B)$ being proportional to $\Xi_0(B)$ similarly to the $T=1$ case. 
In contrast, we show that 
\bea
Z_\alpha^{(0)}-Z_0^{(0)} & = & -(e^{i\alpha}-1)\,\left(\frac{-1}{2\pi}\right)^T
\int_{||B||<\varepsilon} \Bigl(\prod_{t=1}^TdB(t)\Bigr)\,e^{-\frac12 \sum_{t=1}^TB(t)^2} \nn \\
& & \hspace{15mm} \times \int  \Bigl(\prod_{t=1}^Td\phi(t)\Bigr)\,
e^{-i\sum_{t=1}^TB(t)\left\{\phi(t+1)-\phi(t)+W'(\phi(t))\right\}}
\eea 
vanishes as $\varepsilon\to 0$ when $W'(\phi)$ is a polynomial 
(\ref{W'}) with $p\geq 2$.
After the rescaling 
$\phi(t) =\left(\frac{1}{\cN_B}\right)^{\frac{1}{p}}\,\phi'(t)$,    
the $\phi$-integrals become 
\bea
& & \int  \Bigl(\prod_{t=1}^Td\phi(t)\Bigr)\,
e^{-i\sum_{t=1}^TB(t)\left\{\phi(t+1)-\phi(t)+W'(\phi(t))\right\}} \nn \\
& & \hspace{15mm}= \left(\frac{1}{\cN_B}\right)^{\frac{T}{p}}\,\int \Bigl(\prod_{t=1}^Td\phi'(t)\Bigr)\,
e^{-i\sum_{t=1}^T\Omega_B(t) g_p\phi'(t)^p} \left[1+\cO(\varepsilon^{1/p})\right].
\eea
Then, 
\bea
Z_\alpha^{(0)}-Z_0^{(0)} & = & -(e^{i\alpha}-1)\,\left(\frac{-1}{2\pi}\right)^T\,
\left(\int_0^\varepsilon d\cN_B\,\cN_B^{\,T(1-\frac{1}{p})-1}\,e^{-\frac12 \cN_B^2}\right)\,Y  \nn\\
& & \times \left[1+\cO(\varepsilon^{1/p})\right], \\
Y &\equiv &  \int d\Omega_B\int \Bigl(\prod_{t=1}^Td\phi'(t)\Bigr)\,
e^{-i\sum_{t=1}^T\Omega_B(t) g_p\phi'(t)^p}, 
\label{Z_alpha0-Z00_T}
\eea
where the $\cN_B$-integral is $\cO(\varepsilon^{T(1-\frac{1}{p})})$, 
and $Y$ is also shown to be finite in appendix~\ref{app:Y}.   
Thus, we see that $Z_\alpha^{(0)}-Z_0^{(0)}=\cO(\varepsilon^{T(1-\frac{1}{p})})$ vanishes for $p\geq 2$ 
as $\varepsilon$ approaches to zero. 

\paragraph{Unnormalized expectation values}
After the change of variables (\ref{fieldredefT}), the unnormalized expectation values of $B(t)^n$ ($n\geq 1$): 
\be
\vev{B(t)^n}_\alpha'\equiv \int \Bigl(\prod_{t=1}^TdB(t)\Bigr)\,B(t)^n\,\Xi_\alpha(B)
\ee
are expressed as 
\bea
\vev{B(t)^n}_\alpha' & = &  
(e^{i\alpha}-1) \frac{(-1)^{T(T-1)/2}}{(2\pi)^T} \,\int \prod_{t=1}^T\left(dB(t)\,d\tilde{\phi}(t)\right) \,
\frac{B(T)}{\cN_B}\,B(t)^n \nn \\
& & \hspace{21mm}\times \int \Bigl(\prod_{t=2}^T d\psi(t)\Bigr) \left(d\tilde{\bar\psi}\right) \,
e^{-\left.S_0(B, \tilde{\phi}, \psi, \tilde{\bar\psi})\right|_{\psi(1)=0}}. 
\eea  
Here, since the $B$-integrals are not singular for $n\geq 1$, 
we can safely drop trivial Grassmann integrals, differently from the case of 
the partition function $Z_\alpha$.  
This is also the case for more general expectation values  
$\vev{B(1)^{n_1} \cdots B(T)^{n_T}}_\alpha'$ with $n_1,\cdots,n_T\geq 0$ and $\sum_{t=1}^T n_t\geq 1$.

\subsection{Computation of $Y_n(t)$}
\label{app:Yn}
$Y_n(t)$ in (\ref{Yn}) given as 
\bea
Y_n(t) & = & \left(\frac{-1}{2\pi}\right)^T\,\int d\Omega_B\,\Omega_B(t)^n 
\int\prod_{t=1}^T d\phi'(t)\,e^{-i\sum_{t=1}^T\Omega_B(t) g_p\phi'(t)^p} X(\phi'), 
\label{Yn_app}\\
X(\phi') & \equiv & (-1)^{T(T-1)/2}\int \Bigl(\prod_{t=1}^T d\psi(t)\Bigr) \int d\bar\epsilon \int \left(d\bar\psi'\right) 
\,e^{-\sum_{t=1}^T\bar\psi'(t)pg_p\phi'(t)^{p-1}\psi(t)}
\label{X_phi'}
\eea
has trivial Grassmann integrals with respect to $\bar\epsilon$ and one of $\bar\psi'$ which give zero. 
Here, we show that factors in front of these integrals are finite (in fact, they vanish), 
which means that $Y_n(t)$ definitely vanishes. 

First, after $\psi$-integrals, we have 
\be
X(\phi') = (-1)^{T(T-1)/2 +T^2}\left(\prod_{t=1}^T pg_p\phi'(t)^{p-1}\right)\,\int d\bar{\epsilon}
\int\left(d\bar\psi'\right) \prod_{t=1}^T\bar{\psi}'(t). 
\ee
Using 
$\left(d\bar\psi'\right)= 
(-1)^{T-1}\left(\prod_{t=1}^Td\bar{\psi}'(t)\right)\,\left(\sum_{t=1}^T\Omega_B(t)\bar{\psi}'(t)\right)$ 
from the definition (\ref{measure_psibarT}), we obtain 
\be
X(\phi')= (-1)^{T^2}\left(\prod_{t=1}^T pg_p\phi'(t)^{p-1}\right)\,\sum_{t=1}^T\Omega_B(t) \int d\bar{\epsilon}
\left(\int d\bar{\psi}'(t)\bar{\psi}'(t)\bar{\psi}'(t)\right), 
\label{X_phi'_result}
\ee
where the Grassmann integrals $\int d\bar{\epsilon}$ and $\int d\bar{\psi}'(t)\bar{\psi}'(t)\bar{\psi}'(t)$ 
trivially vanish. Thus, (\ref{Yn_app}) has the form 
\bea
Y_n(t) & = & \int d\Omega_B\,\Omega_B(t)^n\,
\prod_{t=1}^T\left(\frac{1}{2\pi}\int^\infty_{-\infty}d\phi'(t) \,e^{-i\Omega_B(t)g_p\phi'(t)^p}\,pg_p\phi'(t)^{p-1}\right) 
\nn \\
 & & \times \sum_{t=1}^T\Omega_B(t) \int d\bar{\epsilon}
\left(\int d\bar{\psi}'(t)\bar{\psi}'(t)\bar{\psi}'(t)\right).
\eea
For the $\phi'(t)$-integral, using the Nicolai mapping, we obtain 
\be
\frac{1}{2\pi}\int^\infty_{-\infty}d\phi'(t) \,e^{-i\Omega_B(t)g_p\phi'(t)^p}\,pg_p\phi'(t)^{p-1}
= \sharp \,\frac{1}{2\pi}\int^\infty_{-\infty} dX(t)\, e^{-i\Omega_B(t) X(t)} = \sharp\,\delta(\Omega_B(t)).
\ee
$\sharp$ is the mapping degree of $X(\phi)=g_p \phi^p$: 
\be
\sharp = \begin{cases} {\rm sgn}(g_p) & \mbox{for $p$: odd} \\
                          0           & \mbox{for $p$: even}, \end{cases}
\label{mapping_deg_app}                          
\ee 
which coincides with the mapping degree of $W'(\phi) = g_p\phi^p + g_{p-1}\phi^{p-1}+\cdots +g_0$. 

As a result,  $Y_n(t)$ has a form  
\be
Y_n(t) = \sharp^T \int d\Omega_B\,\Omega_B(t)^n \left(\prod_{t=1}^T\delta(\Omega_B(t))\right)\,\sum_{t=1}^T 
\Omega_B(t)  \int d\bar{\epsilon}
\left(\int d\bar{\psi}'(t)\bar{\psi}'(t)\bar{\psi}'(t)\right), 
\label{Yn_app2}
\ee
here we note that, since $\left(\Omega_B(1), \cdots,\Omega_B(T)\right)$ is a unit vector 
in ${\bf R}^T$, the integration region for $\Omega_B$ is the unit $(T-1)$-sphere ${\rm S}^{T-1}$ and does not contain the origin, i.e. the support of $\prod_{t=1}^T\delta(\Omega_B(t))$. 
Hence, the $\Omega_B$-integrals (\ref{Yn_app2}) vanish, which shows that $Y_n(t)$ is definitely zero. 

\subsection{Finiteness of $Y$}
\label{app:Y}
In this appendix, we show that 
\be
Y =  \int d\Omega_B\int \Bigl(\prod_{t=1}^Td\phi'(t)\Bigr)\,
e^{-i\sum_{t=1}^T\Omega_B(t) g_p\phi'(t)^p}  
\label{Y_app}
\ee
in (\ref{Z_alpha0-Z00_T}) is finite for $p\geq 2$. 

{}From
\be
\int^\infty_{-\infty}d\phi'\,e^{-ia\phi'^p} 
=\begin{cases}
\frac{2}{|a|^{\frac{1}{p}}} \,e^{-i\,{\rm sgn}(a)\,\frac{\pi}{2p}}\,\Gamma\left(1+\frac{1}{p}\right) & 
(p:\mbox{ even}) \\
\frac{2}{|a|^{\frac{1}{p}}} \,\cos\left(\frac{\pi}{2p}\right)\,\Gamma\left(1+\frac{1}{p}\right) & 
(p:\mbox{ odd}) 
\end{cases}
\ee
for $a\in {\bf R}$, 
we have the bound for $|Y|$: 
\be
|Y|\leq \left(\frac{2}{|g_p|^{\frac{1}{p}}}\,\Gamma\left(1+\frac{1}{p}\right)\right)^T\,
\int d\Omega_B \prod_{t=1}^T |\Omega_B(t)|^{-\frac{1}{p}}.
\label{|Y|_app}
\ee

In the polar coordinates 
\bea
\Omega_B(1) & = & \cos\theta_1, \nn \\
\Omega_B(2) & = & \sin\theta_1\,\cos\theta_2, \nn \\
\vdots      &   &               \nn \\
\Omega_B(T-1) & = & \sin\theta_1\,\sin\theta_2\cdots \sin\theta_{T-2}\,\cos\theta_{T-1}, \nn \\
\Omega_B(T) & = & \sin\theta_1\,\sin\theta_2\cdots \sin\theta_{T-2}\,\sin\theta_{T-1}, 
\eea
with $0\leq \theta_1, \cdots, \theta_{T-2}\leq \pi$ and $0\leq \theta_{T-1}\leq 2\pi$, 
the measure is given by 
\be
d\Omega_B=\sin^{T-2}\theta_1\,\sin^{T-3}\theta_2\,\cdots \sin\theta_{T-2}\,d\theta_1\,d\theta_2\cdots 
d\theta_{T-2}\,d\theta_{T-1}. 
\ee
Then, the $\Omega_B$-integrals in (\ref{|Y|_app}) can be expressed as 
\bea
\int d\Omega_B\,\prod_{t=1}^T |\Omega_B(t)|^{-\frac{1}{p}} & = & 
2^T \prod_{t=1}^{T-1}\left[\int^{\frac{\pi}{2}}_0 d\theta_t \,\left(\cos\theta_t\right)^{-\frac{1}{p}}\,
\left(\sin\theta_t\right)^{(T-t)(1-\frac{1}{p})-1}\right].
\label{int_OmegaB_app}
\eea
For each $t$, since both of the powers of $\cos\theta_t$ and $\sin\theta_t$ are greater than $-1$, 
the $\theta_t$-integral is finite. Thus, (\ref{int_OmegaB_app}) is finite, meaning that $|Y|$ is so.

\section{Computation of $Y_N$}
\label{app:Y_N}
\setcounter{equation}{0}
$Y_N$ given in (\ref{Y_N}), 
\bea
Y_N & \equiv & \left(\frac{-1}{\sqrt{2\pi}}\right)^{N^2} \int d\Omega_B \,\frac{1}{N}\tr\left(\Omega_B^n\right) 
\,\int d^{N^2}\phi' \,e^{-iN\tr (\Omega_B g_p \phi'^p)} \nn \\
 & & \times \int d^{N^2}\psi \,\int d\bar\epsilon \,d^{N^2-1}\bar\psi'\, 
e^{-N\tr\left[\bar\psi' g_p \sum_{\ell=0}^{p-1}\phi'^\ell \psi \phi'^{p-\ell-1}\right]},   
\label{Y_N_app}
\eea 
contains vanishing Grassmann integrals. In this appendix, we explicitly compute $Y_N$ to show that 
prefactors of the vanishing Grassmann integrals are finite. 

In terms of coefficients in the expansion by the basis $\{t^a\}$, the measures are expressed as 
\bea
& & d^{N^2}\phi' = \prod_{a=1}^{N^2} \frac{d\phi'^a}{\sqrt{2\pi}}, \qquad 
d^{N^2}\psi = \prod_{a=1}^{N^2} d\psi^a, \nn\\ 
& & d^{N^2-1}\bar{\psi}' = (-1)^{N^2-1} \left(\prod_{a=1}^{N^2}d\bar\psi'^a\right) \,
\sum_{a=1}^{N^2}\Omega_B^a \bar\psi'^a. 
\eea
After the Grassmann integrals, we obtain 
\bea
Y_N & = & \frac{1}{(2\pi)^{\frac{N^2}{2}}}\,\int d\Omega_B \,\frac{1}{N}\tr (\Omega_B^n)\, 
\left\{\int\Bigl(\prod_{a=1}^{N^2} \frac{d\phi'^a}{\sqrt{2\pi}}\Bigr) \,e^{-i\sum_{a=1}^{N^2} \Omega_B^aV_1(\phi')^a} 
\det_{a,b}\left(V_2(\phi')^{ab}\right) \right\}\nn \\
& & \times \sum_{a=1}^{N^2} \Omega_B^a\, \int d\bar\epsilon 
\left(\int d\bar\psi'^a\,\bar\psi'^a \,\bar\psi'^a\right), 
\label{Y_N_app2}
\eea  
where 
\bea
V_1(\phi')^a & \equiv & N\tr \left(t^a g_p\phi'^p\right), \nn \\
V_2(\phi')^{ab} & \equiv & N\tr \left(t^a g_p\sum_{\ell=0}^{p-1} \phi'^\ell t^b \phi'^{p-\ell -1}\right) 
=\frac{\der}{\der\phi'^b} V_1(\phi')^a. 
\eea
$V'_1(\phi')$ gives the Nicolai mapping to recast the $\phi'$-integrals to 
$
\int  \Bigl(\prod_{a=1}^{N^2} \frac{dV_1^a}{\sqrt{2\pi}}\Bigr) \,e^{-i\sum_{a=1}^{N^2} \Omega_B^aV_1^a}.  
$
The mapping degree of the map $(\phi'^{a=1},\cdots, \phi'^{a=N^2})\to (V_1^{a=1}, \cdots, V_1^{a=N^2})$ 
seems somewhat complicated. 
In order to get a more explicit form of $Y_N$, 
let us move to the expression of $\phi'$ by eigenvalues and $SU(N)$ angles: 
\be
\phi'=U \begin{pmatrix} \lambda_1 &     &     \\
                                & \ddots &     \\
                                 &       &  \lambda_N \end{pmatrix} U^\dagger, \qquad U\in SU(N). 
\ee 
Then, 
\be
V_2(\phi')^{ab} = g_p \sum_{\ell=0}^{p-1} \sum_{i,j=1}^N \lambda_i^\ell \lambda_j^{p-\ell-1} 
\Psi^a_{ji}\Psi^b_{ij}
\qquad 
\mbox{with} \qquad \Psi^a_{ij}\equiv \sqrt{N} (U^\dagger t^a U)_{ij}.
\ee  
Note that, from the completeness of the basis $\{t^a\}$: 
$\sum_{a=1}^{N^2}(t^a)_{ij}(t^a)_{k\ell}=\frac{1}{N}\delta_{i\ell}\delta_{jk}$, 
$\Psi^a_{ij}$ satisfies 
\be
\sum_{a=1}^{N^2}\Psi^a_{ij}\Psi^a_{k\ell}=\delta_{i\ell}\delta_{jk}. 
\ee
Using this, one can see that each of $\Psi_{ij}^a$ ($i,j=1, \cdots, N$) is an eigenvector of $V_2(\phi')^{ab}$ 
whose corresponding eigenvalue is $g_p\sum_{\ell=0}^{p-1} \lambda_i^\ell \lambda_j^{p-\ell-1}$.  
Hence, 
\bea
\det_{a,b}\left(V_2(\phi')^{ab}\right) & = & 
\prod_{i,j=1}^N \left(g_p\sum_{\ell=0}^{p-1} \lambda_i^\ell \lambda_j^{p-\ell-1}\right) \nn \\
 & = & \prod_{i=1}^N\left(g_pp\lambda_i^{p-1}\right)\,
\prod_{i>j}\left(\frac{g_p\lambda_i^p-g_p\lambda_j^p}{\lambda_i-\lambda_j}\right)^2. 
\eea

The measure of $\phi'$ is expressed in terms of the eigenvalues and the angles as 
\be
d^{N^2}\phi' = \prod_{a=1}^{N^2} \frac{d\phi'^a}{\sqrt{2\pi}} 
= \tilde{C}_N \Bigl(\prod_{i=1}^Nd\lambda_i\Bigr) \,\triangle(\lambda)^2\, dU, 
\ee
where $\triangle(\lambda)=\prod_{i>j}(\lambda_i-\lambda_j)$ is the Vandermonde determinant, 
and $dU$ is the $SU(N)$ Haar measure normalized by $\int dU =1$.   
$\tilde{C}_N$ is a numerical constant depending only on $N$. 
Then, the $\phi'$-integrals in (\ref{Y_N_app2}) becomes 
\bea
& & \int\Bigl(\prod_{a=1}^{N^2} \frac{d\phi'^a}{\sqrt{2\pi}}\Bigr) \,
e^{-i\sum_{a=1}^{N^2} \Omega_B^aV_1(\phi')^a}
\nn \\
& & = \tilde{C}_N \int dU \,\int \Bigl(\prod_{i=1}^Nd\lambda_i\Bigr) \,
\prod_{i=1}^N\frac{dw_i}{d\lambda_i}\, \prod_{i>j}(w_i-w_j)^2 \,
e^{-iN\sum_{i=1}^{N} (U^\dagger \Omega_B U)_{ii} w_i},
\eea 
where $w_i\equiv g_p\lambda_i^p$ gives the Nicolai mapping. Thus we find  
\bea
 & & \int\Bigl(\prod_{a=1}^{N^2} \frac{d\phi'^a}{\sqrt{2\pi}}\Bigr) \,
e^{-i\sum_{a=1}^{N^2} \Omega_B^aV_1(\phi')^a}
\nn \\
& & = (-1)^{\frac{N(N-1)}{2}}\,\frac{(2\pi \,\sharp)^N}{N^{N^2}}\, \tilde{C}_N\,
\int dU\, \prod_{i>j} 
\left(\frac{\der}{\der (U^\dagger \Omega_B U)_{ii}} - \frac{\der}{\der (U^\dagger \Omega_B U)_{jj}}\right)^2 
\nn \\
 & & \hspace{56mm} \times \prod_{i=1}^N\delta\left((U^\dagger \Omega_B U)_{ii}\right). 
\eea
$\sharp$ is the mapping degree defined by (\ref{mapping_deg_app}). Plugging this into (\ref{Y_N_app2}), 
we end up with 
\bea
Y_N & = & (-1)^{\frac{N(N-1)}{2}}\,\frac{\sharp^N\, \tilde{C}_N}{(2\pi)^{\frac{N^2}{2}-N}N^{N^2}}\, \int 
d\Omega_B'\, \frac{1}{N}\tr (\Omega_B'^n)\nn \\ 
& & \times \left\{\prod_{i>j}\left(\frac{\der}{\der (\Omega'_B)_{ii}} - \frac{\der}{\der (\Omega'_B)_{jj}}\right)^2 
 \prod_{i=1}^N\delta\left((\Omega_B')_{ii}\right) \right\} \nn \\
& & \times \sum_{a=1}^{N^2} \int dU\, (U\Omega_B' U^\dagger)^a \,
\left(\int d\bar\epsilon\,\int d\bar\psi'^a \, \bar\psi'^a\,\bar\psi'^a\right).
\label{Y_N_app3}
\eea
We changed the integration variable $\Omega_B$ as $\Omega_B'=U^\dagger \Omega_B U$
under which the measure is invariant: $d\Omega_B=d\Omega_B'$. 

In (\ref{Y_N_app3}), the $U$-integrals are clearly finite. The integration region for  
$\Omega_B'$ is the $(N^2-1)$-sphere ${\rm S}^{N^2-1}$ defined by 
$\sum_{i,j=1}^{N} |(\Omega_B')_{ij}|^2=\frac{1}{N}$, and the support of the delta-function 
$\prod_{i=1}^N\delta\left((\Omega_B')_{ii}\right)$ is a region of the ${\rm S}^{N^2-1}$ determined by 
$(\Omega_B')_{11}=\cdots =(\Omega_B')_{NN}=0$, i.e. ${\rm S}^{N^2-N-1}$. 
The $\Omega_B'$-integrals are also finite, because the integrand is a polynomial of $(\Omega_B')_{ij}$ 
multiplied by the delta-functions. 
Thus, the prefactors of the vanishing Grassmann integrals in (\ref{Y_N_app3}) are finite, 
meaning that $Y_N$ definitely vanishes due to the trivial Grassmann integrals.



\begin{thebibliography}{999}
{\small 
\bibitem{Witten:1981nf}
  E.~Witten,
  ``Dynamical Breaking Of Supersymmetry,''
  Nucl.\ Phys.\  B {\bf 188} (1981) 513.

  
\bibitem{Banks:1996vh}
  T.~Banks, W.~Fischler, S.~H.~Shenker and L.~Susskind,
  ``M theory as a matrix model: A conjecture,''
  Phys.\ Rev.\  D {\bf 55} (1997) 5112
  [{\tt arXiv:hep-th/9610043}].

\bibitem{Ishibashi:1996xs}
N.~Ishibashi, H.~Kawai, Y.~Kitazawa and A.~Tsuchiya,
 ``A large-N reduced model as superstring,''
  Nucl.\ Phys.\  B {\bf 498} (1997) 467
  [{\tt arXiv:hep-th/9612115}].
  
\bibitem{Dijkgraaf:1997vv}
  R.~Dijkgraaf, E.~P.~Verlinde and H.~L.~Verlinde,
  ``Matrix string theory,''
  Nucl.\ Phys.\  B {\bf 500} (1997) 43
  [{\tt arXiv:hep-th/9703030}].
  
\bibitem{Nishimura:2001sx}
  J.~Nishimura and F.~Sugino,
  ``Dynamical generation of four-dimensional space-time in the IIB matrix model,''
  JHEP {\bf 0205} (2002) 001
  [{\tt arXiv:hep-th/0111102}].\\
  H.~Kawai, S.~Kawamoto, T.~Kuroki, T.~Matsuo and S.~Shinohara,
 ``Mean field approximation of IIB matrix model and emergence of four dimensional space-time,''
  Nucl.\ Phys.\  B {\bf 647} (2002) 153
  [{\tt arXiv:hep-th/0204240}]. \\
H.~Kawai, S.~Kawamoto, T.~Kuroki and S.~Shinohara,
 ``Improved perturbation theory and four-dimensional space-time in IIB matrix model,''
  Prog.\ Theor.\ Phys.\  {\bf 109} (2003) 115
  [{\tt arXiv:hep-th/0211272}]. \\
T.~Aoyama, H.~Kawai and Y.~Shibusa,
``Stability of 4-dimensional space-time from IIB matrix model via improved mean field approximation,''
  Prog.\ Theor.\ Phys.\  {\bf 115} (2006) 1179
  [{\tt arXiv:hep-th/0602244}]. \\
T.~Aoyama and H.~Kawai,
``Higher order terms of improved mean field approximation for IIB matrix model and emergence of four-dimensional space-time,''
  Prog.\ Theor.\ Phys.\  {\bf 116} (2006) 405
  [{\tt arXiv:hep-th/0603146}]. \\
T.~Aoyama and Y.~Shibusa,
 ``Improved perturbation method and its application to the IIB matrix model,''
  Nucl.\ Phys.\  B {\bf 754} (2006) 48
  [{\tt arXiv:hep-th/0604211}].

\bibitem{Kuroki:2009yg}
  T.~Kuroki and F.~Sugino,
  ``Spontaneous supersymmetry breaking in large-$N$ matrix models with slowly
  varying potential,''
  Nucl.\ Phys.\  B {\bf 830} (2010) 434 
  [{\tt arXiv:0909.3952 [hep-th]}].

\bibitem{Witten:1982df}
  E.~Witten,
  ``Constraints On Supersymmetry Breaking,''
  Nucl.\ Phys.\  B {\bf 202} (1982) 253.


\bibitem{Witten:1991mk}
  E.~Witten,
  ``The N Matrix Model And Gauged WZW Models,''
  Nucl.\ Phys.\  B {\bf 371} (1992) 191.

\bibitem{Nicolai:1979nr}
  H.~Nicolai,
  ``On A New Characterization Of Scalar Supersymmetric Theories,''
  Phys.\ Lett.\  B {\bf 89} (1980) 341;
  ``SUPERSYMMETRY AND FUNCTIONAL INTEGRATION MEASURES,''
  Nucl.\ Phys.\  B {\bf 176} (1980) 419.





\bibitem{Hori:2003ic}
  K.~Hori {\it et al.},
  ``Mirror symmetry,''
{\it  Providence, USA: AMS (2003) 929 p}.


\bibitem{Marinari:1990jc}
  E.~Marinari and G.~Parisi,
 ``THE SUPERSYMMETRIC ONE-DIMENSIONAL STRING,''
  Phys.\ Lett.\  B {\bf 240} (1990) 375.


\bibitem{Nojiri:1992zu}
  S.~Nojiri,
  ``$N = \frac{1}{2}$ superstring in zero-dimension and the spontaneous breakdown of the supersymmetry,''
  Mod.\ Phys.\ Lett.\  A {\bf 7} (1992) 2979
  [{\tt arXiv:hep-th/9206086}].


\bibitem{Eynard:1995nv}
  B.~Eynard and C.~Kristjansen,
  ``Exact Solution of the $O(n)$ Model on a Random Lattice,''
  Nucl.\ Phys.\  B {\bf 455} (1995) 577
  [{\tt arXiv:hep-th/9506193}].  
 
 
\bibitem{Sakai:1983dg}
  N.~Sakai and M.~Sakamoto,
  ``Lattice Supersymmetry And The Nicolai Mapping,''
  Nucl.\ Phys.\  B {\bf 229} (1983) 173.

  Y.~Kikukawa and Y.~Nakayama,
  ``Nicolai mapping vs. exact chiral symmetry on the lattice,''
  Phys.\ Rev.\  D {\bf 66} (2002) 094508
  [{\tt arXiv:hep-lat/0207013}].
   
\bibitem{Kaplan_review}
  S.~Catterall, D.~B.~Kaplan and M.~\"Unsal,
  ``Exact lattice supersymmetry,''
  Phys.\ Rept.\  {\bf 484} (2009) 71
  [{\tt arXiv:0903.4881 [hep-lat]}].
  
\bibitem{Kaplan:2002wv}
  D.~B.~Kaplan, E.~Katz and M.~\"Unsal,
  ``Supersymmetry on a spatial lattice,''
  JHEP {\bf 0305} (2003) 037
  [{\tt arXiv:hep-lat/0206019}].

  A.~G.~Cohen, D.~B.~Kaplan, E.~Katz and M.~\"Unsal,
  ``Supersymmetry on a Euclidean spacetime lattice. I: A target theory with four supercharges,''
  JHEP {\bf 0308} (2003) 024
  [{\tt arXiv:hep-lat/0302017}]; 
%
  ``Supersymmetry on a Euclidean spacetime lattice. II: Target theories with eight supercharges,''
  JHEP {\bf 0312} (2003) 031
  [{\tt arXiv:hep-lat/0307012}].

  D.~B.~Kaplan and M.~\"Unsal,
  ``A Euclidean lattice construction of supersymmetric Yang-Mills theories with sixteen supercharges,''
  JHEP {\bf 0509} (2005) 042
  [{\tt arXiv:hep-lat/0503039}].

\bibitem{Catterall:2003wd}
  S.~Catterall,
  ``Lattice Supersymmetry and Topological Field Theory,''
  JHEP {\bf 0305} (2003) 038
  [{\tt arXiv:hep-lat/0301028}].

  S.~Catterall and S.~Ghadab,
  ``Lattice sigma models with exact supersymmetry,''
  JHEP {\bf 0405} (2004) 044
  [{\tt arXiv:hep-lat/0311042}]; 
%
  ``Twisted supersymmetric sigma model on the lattice,''
  JHEP {\bf 0610} (2006) 063
  [{\tt arXiv:hep-lat/0607010}].

  

\bibitem{Catterall:2004np}
  S.~Catterall,
  ``A geometrical approach to N = 2 super Yang-Mills theory on the two dimensional lattice,''
  JHEP {\bf 0411} (2004) 006
  [{\tt arXiv:hep-lat/0410052}]; 
%
  ``Lattice formulation of N = 4 super Yang-Mills theory,''
  JHEP {\bf 0506} (2005) 027
  [{\tt arXiv:hep-lat/0503036}].

  

\bibitem{SYM_lat}
  F.~Sugino,
  ``Super Yang-Mills theories on the two-dimensional lattice with exact supersymmetry,''
  JHEP {\bf 0403} (2004) 067
  [{\tt arXiv:hep-lat/0401017}];
%
 ``Two-dimensional compact N = (2,2) lattice super Yang-Mills theory with exact supersymmetry,''
  Phys.\ Lett.\  B {\bf 635} (2006) 218
  [{\tt arXiv:hep-lat/0601024}].

\bibitem{Endres:2006ic}
  M.~G.~Endres and D.~B.~Kaplan,
  ``Lattice formulation of (2,2) supersymmetric gauge theories with matter fields,''
  JHEP {\bf 0610} (2006) 076
  [{\tt arXiv:hep-lat/0604012}]. 

  S.~Matsuura,
  ``Two-dimensional N=(2,2) Supersymmetric Lattice Gauge Theory with Matter Fields in the Fundamental Representation,''
  JHEP {\bf 0807} (2008) 127
  [{\tt arXiv:0805.4491 [hep-th]}].


\bibitem{SQCD_lat}
 F.~Sugino,
  ``Lattice Formulation of Two-Dimensional N=(2,2) SQCD with Exact Supersymmetry,''
  Nucl.\ Phys.\  B {\bf 808} (2009) 292
  [{\tt arXiv:0807.2683 [hep-lat]}].

  Y.~Kikukawa and F.~Sugino,
  ``Ginsparg-Wilson Formulation of 2D N =(2,2) SQCD with Exact Lattice Supersymmetry,''
  Nucl.\ Phys.\  B {\bf 819} (2009) 76
  [{\tt arXiv:0811.0916 [hep-lat]}].

  D.~Kadoh, F.~Sugino and H.~Suzuki,
  ``Lattice formulation of 2D $\mathcal{N}=(2,2)$ SQCD based on the B model twist,''
  Nucl.\ Phys.\  B {\bf 820} (2009) 99
  [{\tt arXiv:0903.5398 [hep-lat]}].

\bibitem{Ishii:2008ib}
  T.~Ishii, G.~Ishiki, S.~Shimasaki and A.~Tsuchiya,
  ``N=4 Super Yang-Mills from the Plane Wave Matrix Model,''
  Phys.\ Rev.\  D {\bf 78} (2008) 106001
  [{\tt arXiv:0807.2352 [hep-th]}].

\bibitem{Hanada:2010kt}
  M.~Hanada, S.~Matsuura and F.~Sugino,
  ``Two-dimensional lattice for four-dimensional N=4 supersymmetric Yang-Mills,''
  {\tt arXiv:1004.5513 [hep-lat]}.

  M.~Hanada,
  ``A fine tuning free formulation of 4d N=4 super Yang-Mills,''
  {\tt arXiv:1009.0901 [hep-lat]}.

\bibitem{D'Adda:2010pg}
  A.~D'Adda, A.~Feo, I.~Kanamori, N.~Kawamoto and J.~Saito,
  ``Species Doublers as Super Multiplets in Lattice Supersymmetry: Exact Supersymmetry with Interactions for D=1 N=2,''
  JHEP {\bf 1009} (2010) 059
  [{\tt arXiv:1006.2046 [hep-lat]}].




 
\bibitem{Kuroki:2007iy}
  T.~Kuroki and F.~Sugino,
  ``Spontaneous Supersymmetry Breaking by Large-N Matrices,''
  Nucl.\ Phys.\  B {\bf 796} (2008) 471
  [{\tt arXiv:0710.3971 [hep-th]}].
  
\bibitem{Oda:2000im}
  S.~Oda and F.~Sugino,
  ``Gaussian and mean field approximations for reduced Yang-Mills integrals,''
  JHEP {\bf 0103} (2001) 026
  [{\tt arXiv:hep-th/0011175}].\\
  F.~Sugino,
  ``Gaussian and mean field approximations for reduced 4D supersymmetric Yang-Mills integral,''
  JHEP {\bf 0107} (2001) 014
  [{\tt arXiv:hep-th/0105284}].
  
\bibitem{Nishimura:2002va}
  J.~Nishimura, T.~Okubo and F.~Sugino,
  ``Convergent Gaussian expansion method: Demonstration in reduced Yang-Mills integrals,''
  JHEP {\bf 0210} (2002) 043
  [{\tt arXiv:hep-th/0205253}];
  ``Testing the Gaussian expansion method in exactly solvable matrix models,''
  JHEP {\bf 0310} (2003) 057
  [{\tt arXiv:hep-th/0309262}].


\bibitem{Aoyama:2006yx}
  T.~Aoyama, T.~Kuroki and Y.~Shibusa,
  ``Dynamical generation of non-Abelian gauge group via the improved perturbation theory,''
  Phys.\ Rev.\  D {\bf 74} (2006) 106004
  [{\tt arXiv:hep-th/0608031}].
  

\bibitem{Aoyama:2010ry}
  T.~Aoyama, J.~Nishimura and T.~Okubo,
  ``Spontaneous breaking of the rotational symmetry in dimensionally reduced super Yang-Mills models,''
  {\tt arXiv:1007.0883 [hep-th]}.

\bibitem{Giedt:2004qs}
  J.~Giedt and E.~Poppitz,
  ``Lattice supersymmetry, superfields and renormalization,''
  JHEP {\bf 0409} (2004) 029
  [{\tt arXiv:hep-th/0407135}].






      
}

\end{thebibliography}
\end{document}